\documentclass[aps,prb,twocolumn,nopacs,amssymb]{revtex4}

\usepackage{amsmath}
\usepackage[dvips]{graphicx}
\usepackage{dcolumn}
\usepackage{bm}
\def\be{\begin{equation}}
\def\en{\end{equation}}

\def\p{\partial} 
\newcommand{\av}[1]{\langle{#1}\rangle}
\newcommand{\AV}[1]{{\bigg\langle}{#1}{\bigg\rangle}}
\def\gs{\gtrsim}
\def\ls{\lesssim}
\newcommand{\bi}[1]{\mbox{\boldmath$#1$}}
\def\p{\partial}
\def\bea{\begin{eqnarray}}
\def\ena{\end{eqnarray}}

\newcommand{\ppp}[3]{{\bigg(}\frac{\partial {#1}}{\partial {#2}}{\bigg )}_{#3}}

\def\asigma{\stackrel{\leftrightarrow}{\sigma}}

\def\he{{\hat{e}}} 
\def\hn{{\hat{n}}} 
\def\hQ{{\hat{q}}}

\begin{document}


\title{Theory of Applying 
Heat Flow from   Thermostatted  Boundary Walls:
\\ Dissipative and 
 Local-Equilibrium  Responses 
and Fluctuation Theorems
}

\author{ Akira Onuki }
\affiliation{
Department of Physics, Kyoto University, Kyoto 606-8502, 
Japan\footnote{email:onuki@scphys.kyoto-u.ac.jp} 
}

\date{\today}

\begin{abstract}
We construct a microscopic  theory of 
 applying a heat flow  from thermostatted 
boundary walls in the film geometry. 
We treat a classical one-component fluid, 
but  our method is applicable to any fluids and  solids. 
We express linear response of any variable ${\cal B}$  
in terms of the time-correlation functions 
between $\cal B$ and  
the  heat flows ${\cal J}_K$ 
from  the thermostats to the particles. 
Furthermore, the surface variables 
 ${\cal J}_K$  can be written in the form of 
 space integrals of  bulk quantities 
from  the equations of motion. Owing  to this surface-to-bulk 
relation,  the steady-state response functions  consist  of dissipative  
and  local-equilibrium parts, where the former gives rise to 
Fourier's law with Green's expression for the thermal conductivity. 
In the nonlinear regime, we  derive 
  the steady-state distribution 
in the phase space in the  McLennan-Zubarev form from the first principles. 
  Some  fluctuation theorems are also presented. 
\end{abstract} 


  

\maketitle

\section{Introduction}

In the linear response theory  in statistical 
mechanics, response of any physical quantity to a small 
perturbation is   expressed  in terms of appropriate 
time-correlation functions\cite{Mor,Hansen,Onukibook,Zwan}. 
For thermal disturbances,  
Green\cite{Green}   expressed  the transport coefficients 
 such as  the viscosities and the thermal conductivity 
in terms of  the time-correlation functions of the stress 
and the heat flux, respectively.   These expressions    also followed   
from the relaxations   of the 
time-correlation functions  of  the  hydrodynamic 
variables\cite{Hansen,Kada,Zwan}. 
These {\it gross variables} 
  obey the fluctuating 
hydrodynamic equations with random stress and heat current 
slightly away from equilibrium\cite{Landau-f}, where 
the latter satisfy the fluctuation-dissipation 
relations\cite{Onsager,Kubo1,Mori,Zwanzig1961}.
Some attempts were also made to develop  nonlinear 
theories of thermal disturbances\cite{Mori1,McL,Zu,Gunton}.  
On the other hand, for externally applied 
 fields  (such as magnetic field), 
   Kubo\cite{Kubo}  developed a  
linear response theory, where 
  the  Hamiltonian consists  of the unperturbed  one  $\cal H$ 
and  a small time-dependent perturbation  as  
$
{\cal H}'= {\cal H}- \gamma_{\rm ex} (t) {\cal A}. 
$   
Here,   $\gamma_{\rm ex}(t)$ is  an   applied field  
 and ${\cal A}$ is its  conjugate variable.

Thermal disturbances are usually induced from   
 boundaries, while  the early 
 theories\cite{Onsager,Green,Mori1,McL,Zu,Gunton,Oppen}  
  started with  nearly  homogeneous  
velocity and temperature  gradients   in the cell interior 
and are not  directly applicable to heterogeneous systems. 
Hence,   we should  develop a  theory 
treating  perturbations applied at   boundaries. 
With this objective, we recently applied shear strains 
 to   particle systems by slightly 
moving  boundary walls\cite{Kawasaki}. 
  We   set   up a Hamiltonian in Kubo's  form, 
where  the  applied  field  $\gamma_{\rm ex}(t)$  
is  the  {\it mean}  shear strain  in a film 
 and the  conjugate variable 
$\cal A$ is expressed in terms of  
 the  tangential forces from the bottom and top 
walls to the particles.   
Then,  linear  response of  any variable $\cal B$  can be expressed 
 in terms of the time-correlation function 
of ${\cal B}$ and $ {\cal A}$.  As a key relation,  we further set 
 ${\cal A}=\int d{\bi r}[ \Pi_{xz}- z \p J_x/\p t]$, 
where 
 $ \Pi_{xz}$ is the shear stress ($xz$ component), and $J_x$ 
is the momentum density ($x$ component) in the bulk. 
  This {\it surface-to-bulk}  relation 
 reveals  how the surface forces induce  bulk response, leading  to 
 Green's expression  for the shear viscosity $\eta$. 
It is worth noting that  Petravic and Harrowell\cite{Harrowell} 
expressed   $\eta$ 
in terms of   the surface-force time-correlation.

The thermal conductivity  $\lambda$  (as well as $\eta$)  can  be calculated  
  from  the    Green-Kubo  formula via 
 equilibrium  molecular dynamics simulation\cite{Mor,Hansen}. 
However, to  study  nonequilibrium steady  states, 
we need at least two  thermostats 
at different temperatures\cite{Bone}.  There have been 
 a number of such nonequilibrium simulations. In particular, a  heat 
flow was applied  across solid-liquid\cite{Harrowell,Barrat}  
and  gas-liquid\cite{Hafs} interfaces, in 
near-critical fluids\cite{Hamanaka}, and 
in one-dimensional lattices\cite{Lepri,Dhar,Sano}.
We also mention  the   theories on 
fluctuation theorems\cite{FD,Jar,Jar1,Cr,Lebo2,Co,Evans1,Ri,Se}. 
In  their application to  heat-conducting steady 
states\cite{Ko,Ja2,Searles1,W,W1,Ha}, 
  relevant are  the 
  heat flows  ${\cal J}_K$  from the thermostats  to 
the particles, where $K$ represents the  {\it top}  
or {\it bottom}  thermostat in the film geometry.   
Then, in the linear regime,   $\lambda$ can be expressed 
 in terms of   the  surface variables  ${\cal J}_{K}$.
However,  Green's   expression for 
$\lambda$ is written  in terms of the 
heat flux in the bulk\cite{Harrowell,Bone}, so it  has  not been derived  from 
the fluctuation theorems.
On the other hand, the Green-Kubo formula readily follows 
from  the fluctuation theorems for 
 a  perturbation  applied in the bulk region\cite{FD,Co,Evans1,Lebo2}.

In this paper, we use  Nos\'e-Hoover 
thermostats\cite{Hansen,Nose,Hoover,Mor} 
in top and bottom boundary layers 
 in contact with a classical fluid. 
We induce heat conduction by changing the 
boundary temperatures 
  by $\delta T_K(t)$ for $t>0$.  
 In the linear regime, we obtain  
 response relations of any variable $\cal B$ 
in terms of the time-correlation functions  
$\av{{\cal B}(t) {\cal J}_K(0)}_{\rm e}$ in equilibrium.  
As in the case of boundary-driven  strains\cite{Kawasaki}, 
we   further express  ${\cal J}_K$ 
 as a sum  of    bulk terms, where 
one  term  yields   local-equilibrium response 
and another one    dissipative response. 
Remarkably, the latter  
is proportional to the   heat flux integral 
 $\int d{\bi r} J_h^z({\bi r},t)$,   giving 
 rise to    Green's expression for $\lambda$. 
We then  calculate the relaxations of  
the hydrodynamic variables   after a boundary temperature 
change, where we account for  the adiabatic (piston) effect at a  fixed cell 
volume\cite{Za,Ferrell,Miura,Onukibook}. 
These average deviations are  related to 
 $\av{{\cal B}(t) {\cal J}_K(0)}_{\rm e}$.  
We also calculate  the Evans-Searles 
dissipation function\cite{Evans1,Searles1} analytically 
in transient states in the linear regime, which 
is related to the entropy production and the logarithm 
of the distribution function of the hydrodynamic variables.

In this paper, we also study   nonlinear dynamics   in the phase space 
 on the basis of  the research on  fluctuation 
theorems\cite{FD,Se,Ha,Ri,Jar,Jar1,Ja2,Cr,Lebo2,Co,Evans1,Ko,Searles1,W,W1},  
where   the boundary temperature changes 
$\delta T_K$ need not be small.  
It enables us to 
 understand the linear theory itself from a wide perspective.  
  We  add    some  nonlinear 
results particularly in   transient states 
 for  $\delta T_{\rm top}= 
\delta T_{\rm bot}$ and in   steady states  
 for $\delta T_{\rm top}\neq \delta T_{\rm top}$.  
In the latter finding, 
we obtain   the  steady-state  distribution 
in   the classical 
form of  McLennan\cite{McL} and Zubarev\cite{Zu} 
from  the first principles. 
Here,  using  the surface-to-bulk relationship, 
  we can introduce  the   local-equilibrium 
distribution with a space-dependent   temperature, from which  
relaxation to a  steady state occurs 
in a microscopic time.

The organization of this paper is as follows. In Sec. II, we
will  present the theoretical background  of 
our thermostatted system.    In Sec.III, 
the linear response  will be  discussed 
using the  Liouville equation. 
In Sec.IV, a nonlinear theory will be presented. 
We will also  calculate the relaxation 
behaviors 
 of the hydrodynamic variables in Appendix A and 
 the dissipation function  in Appendix B  
after a boundary temperature change. 
A theory of  long-range correlations  
in the canonical ensemble\cite{Lebo} 
will be presented in Appendix C, which 
is crucial in calculating the local-equilibrium 
response.

\section{Theoretical background   }

We treat a classical one-component  fluid    
with   particle number $N$ in a film cell. 
We write the particle 
positions  as  ${\bi r}_i=(x_i,y_i,z_i)$,   the  velocities   as 
$ {\bi v}_i = {\dot{\bi r}}_i=(v_i^x,v_i^y,v_i^z)$, and the 
momenta as ${{\bi p}}_i= m{\bi v}_i =(p_i^x,p_i^y,p_i^z)$ 
with $m$ being  the particle  mass. 
Herafter, the upper dot denotes taking the  time derivative $\p/\p t$. 
The space dimensionality $d$ can also be  one or  two. 
We write the cell   thickness as $H$ and the  lateral dimension as $L$ 
in the $xy$ plane.   The surface area of each boundary layer 
is $A= L^{d-1}$ and the cell volume is  
  $V =HA$.  We can impose  the periodic boundary condition  
 along the $x$ and $y$ axes.  These lengths  are   much 
longer than the particle diameters.  The fluid is 
in a liquid or supercritical one-phase state 
with a high density $n=N/V$.

{\subsection{Heat variable and heat inputs}}

We  introduce    thermostatted 
 boundary layers\cite{Shiba}, which are  in the  regions 
$-\ell_{\rm w}<z<0$ at the bottom and $0<z-H<\ell_{\rm w}$ at the top. 
To each layer, $M$  particles are bound   
by spring  potentials $\psi (|{\bi r}_k -  {\bi R}_k|)
$, where    ${\bi r}_k$ are the positions of the bound  particles and 
   ${\bi R}_k$  are the  pinning centers fixed to  the layers at a 
high density.  We  assume  $\ell_{\rm w}\ll H$ and $M\ll N$ for simplicity.
In our previous papers\cite{Kawasaki,Shiba}, 
 we used the  harmonic form  $\psi(r)=s_0 r^2/2$, where    impenetrable 
walls were realized for large $s_0$. In one-dimensional model 
solids\cite{Lepri,Dhar,Sano}, only 
the end particles were thermostatted (where   $M=1$).

Particle pairs $i$ and $j$   
interact  via short-range    potentials 
$\phi_{ab} (r_{ij}) $ with    $r_{ij}= |{\bi r}_i-{\bi r}_j|$, where 
$i\in a$ and $j \in b$ with  
$a$ and $b$ denoting  either of unbound or bound particles.  
 We write $\phi_{ij}= 
\phi_{ab} (r_{ij}) $ and  $\psi_{k}= 
\psi (|{\bi r}_k -  {\bi R}_k|) $.  The  
total  energy  of the particles is  written as    
\be 
 {\cal H}= \sum_i \frac{1}{2m_i} |{\bi p}_i|^2 + 
\frac{1}{2} \sum_{i,j} \phi_{ij}+ \sum_{k>N}  \psi_{k} 
\en 
where we sum over all the particles.
The mass of the bound particles  
can  differ from that of the unbound ones.

We define  the number   density  and 
the  energy  density of the unbound particles microscopically 
as 
\bea
&& \hn({\bi r},t) =\sum_{i\le N}   \delta({\bi r}- {\bi r}_i), 
\\
&& \he({\bi r},t) =\sum_{i\le N} e_i  \delta
({\bi r}- {\bi r}_i),
\ena 
 where  $ 
e_i= m |{\bi v}_i|^2/2+ \sum_{ j\le N}\phi_{ij}/2$ 
is the energy supported by  particle $i$.      Hereafter, 
the variables with an upper caret symbol are 
microscopically defined space-dependent  variables.  
Following the literature\cite{Hansen,Onukibook,Kada}, 
we   introduce the heat variable $\hQ({\bi r},t)$ by 
\be 
\hQ({\bi r},t)=\he({\bi r},t)-h \hn({\bi r},t). 
\en 
Here,  $h= (e+p)/n$ is the enthalpy per particle, 
where $e$, $p$, and $n$ are  
the thermodynamic  energy density,  pressure,  and 
 number density, respectively.  
Then, the equilibrium average of $\hQ$ is equal to $-p$ 
and the entropy per particle $s$ 
satisfies  the thermodynamic differential 
relation $nT ds= de-hdn$, so we can treat 
$ \delta \hQ = \hQ+p  
$ as   the deviation of the  fluctuating 
 entropy variable multiplied by $nT$. 
See   the last paragraph 
of this subsection for more discussions on 
the flux of $\hQ$.

The unbound particles ($i\le N$) obey 
 Newton's  equations   $m {\dot{\bi v}}_i =  {\bi f}_i$ 
with  ${\bi f}_i=  -\p {\cal H}/\p {\bi r}_i$ 
being  the force on  particle $i$. 
Then,  ${ \hQ}$ evolves in time as  
\be 
\frac{\p}{\p t}{ \hQ}+\nabla\cdot {\bi J}_h  = \sum_{i\le N} 
{\bi v}_i\cdot ({\bi f}_i^{{\rm top}}+{\bi f}_i^{{\rm bot}})
   \delta({\bi r}- {\bi r}_i).
\en 
where  ${\bi J}_h({\bi r},t)$ is the  heat flux in the cell 
and  the right hand side represents  the heat-input density  from the walls. 
 We write the forces  on unbound  particle  $i\le  N$ 
from the top and bottom bound particles $(k>N)$ as 
\be 
{\bi f}^{{\rm top}}_i    
 =  - \sum_{k\in {\rm top}} {\bold\nabla}_i  \phi_{ik},\quad     
{\bi f}^{{\rm bot}}_i   
 =  - \sum_{k\in {\rm bot}} {\bold\nabla}_i  \phi_{ik},
\en  
where  $\nabla_i=\p/\p {\bi r}_i=(\nabla_i^x, \nabla_i^y, \nabla_i^z)$.
  The  $\alpha$ component of ${\bi J}_h$ ($\alpha=x,y, z$)  is microscopically 
 expressed as  
\bea
&&\hspace{-8mm}
{ J}^\alpha_h =
\sum_{i\le N}(e_i - h){v}_i^\alpha \delta({\bi r}-{\bi r}_i)\nonumber\\
&&\hspace{-4mm} -   \sum_{i,j\le N } \frac{1}
{2r_{ij}} \phi_{ij}'   x_{ij}^\alpha ({\bi r}_{ij}\cdot {\bi v}_j)   
 {\hat{\delta}} ({\bi r},{\bi r}_i,{\bi r}_j) ,
\ena  
where $\phi_{ij}'={d\phi_{ij}}/{dr_{ij}}$ and 
$x_{ij}^\alpha $ is  the $\alpha $ 
component of ${\bi r}_{ij}= {\bi r}_i-{\bi r}_j$. 
We use   the Irving-Kirkwood delta    function\cite{Irving}    
 $ {\hat{\delta}} ({\bi r},{\bi r}_i,{\bi r}_j)= 
 \int_0^1 d\lambda {\delta}({\bi r}-\lambda{\bi r}_i -(1-\lambda){\bi r}_j)$, 
which is nonvanishing on the line segment connecting ${\bi r}_i$ 
and ${\bi r}_j$.  The microscopic expression for the  pressure is then   
\be 
{\hat p}= 
\sum_{i\le N} \frac{|{\bi p}_i|^2}{dm}  \delta({\bi r}-{\bi r}_i)
- \hspace{-1mm}
 \sum_{i,j\le N }\frac{r_{ij}}{2d}  \phi_{ij}' 
 {\hat{\delta}} ({\bi r},{\bi r}_i,{\bi r}_j).
\en 
where the equilibrium average $p(T)= \av{{\hat p}}_{\rm e}$ 
in the bulk region is the 
the thermodynamic pressure.  

It is convenient to 
 define the total  internal     energy of the unbound particles as 
\be 
{\cal H}_{\rm in}= \sum_{i\le N}e_i
=  \sum_{i\le N}\frac{1}{2m} |{\bi p}_i|^2
+\frac{1}{2} \sum_{i\le N,j\le N} \phi_{ij}. 
\en 
At fixed  volume $V$ and  $N$, ${\cal H}_{\rm in}$ changes  in time as  
\be 
{\dot{\cal H}}_{\rm in}= \frac{d}{dt}\int d{\bi r} {\hQ}= {\cal I}_{\rm top} + 
{\cal I}_{\rm bot}, 
\en 
where we use $\int d{\bi r} { \hn}=N$ and the 
space integral $\int d{\bi r}(\cdots)$ is taken in a region containing 
all the unbound particles. 
The ${\cal I}_K$ is  the  heat flow   
from the bound particles in the layer $K$ to the unbound particles 
defined as  
\be 
{\cal I}_K = \sum_{i\le N} {\bi v}_i\cdot {\bi f}_i^{K}. 
\en 
Hereafter, the subscript  $K$ represents  
top or bottom.

Using Eq.(5) we also  integrate 
 the product $z{\p \hQ}/\p t$ as 
\bea 
\frac{d}{dt} 
\int d{\bi r} z\hQ- \int d{\bi r}J_h^z &=&
\sum_K  \sum_{i\le N}  z_i 
{\bi v}_i\cdot {\bi f}_i^{{\rm K}}\nonumber\\
&=  & H {\cal I}_{\rm top}.  
\ena 
In the first line, the left hand side 
is written  in terms of the degrees of freedom 
of the unbound particles, while the right hand side 
involves the bound ones. 
In the second line, we retain  
 the contribution at the top  replacing  
 $z_i$ by $H$ from $|z_i-H|<\ell_{\rm w}\ll H$.
We thus find  
\be
{\cal I}_{\rm top}= 
\int d{\bi r}\frac{ z}{ H}\frac{\p \hQ}{\p t} 
- \frac{1}{H}{\cal G}, 
 \quad 
{\cal I}_{\rm bot}= {{\dot{\cal H}}}_{\rm in}
- {\cal I}_{\rm top}, 
\en
where we define the integrated heat flux along the $z$ axis,
\be 
{\cal G} (t) = \int d{\bi r} J_h^z({\bi r},t).
\en

In the periodic boundary condition, 
the surface terms are nonexistent in Eq.(12).
Then,  the   space-time integral   of the heat flux  
$\int_0^t ds {\cal G}(s)$ is equal to 
\be 
G_\lambda(t)=  \sum_i z_i (t) [e_i(t)-\av{e_i}_{\rm e}]. 
\en 
This yields  Helfand's  formula\cite{Helfand,Gaspard}
  for the thermal conductivity  
$\lambda= \lim_{t\to \infty} 
 [\av{[G_\lambda(t)- G_\lambda(0)]^2}_{\rm e}
/2tVk_BT^2]$, which 
is equivalent to Green's one in Eq.(45)\cite{Green}.

Between  the heat flux $J_h^\alpha({\bi r})$   and  the momentum density 
$
J^\alpha({\bi r}) = \sum_{i\le N}   p_i^\alpha \delta ({\bi r}- {\bi r}_i),
$     
 we can consider   their  equal-time 
correlation   $G_{\alpha\beta}({\bi r}-{\bi r}', t)
=\av{J_h^\alpha ({\bi r}) J^\beta({\bi r}')}_{\rm e}$. 
It is a short-ranged function of ${\bi r}- {\bi r}'$ 
and its space integral vanishes. 
Namely, we have  $\int d{\bi r}G_{\alpha\beta}({\bi r})=0$. 
 This means that 
 ${ J}^\alpha_h({\bi r})$ is {\it orthogonal} to  
the long-wavelength hydrodynamic 
variables, which is 
crucial  in the projection operator formalism 
of irreversibility\cite{Mori,Zwanzig1961,Gunton}.

{\subsection{ Nos\'e-Hoover thermostats in boundary layers}}

We attach  Nos\'e-Hoover thermostats  
to the bound particles at the top and the bottom 
separately\cite{Hansen,Mor,Nose,Hoover}. 
Their equations of motion are deterministic and are written as 
\be 
 {\dot {\bi p}}_k
= {\bi f}_k  - \zeta_K   {\bi p}_k \quad (k>N), 
\en 
where  ${\bi p}_k =m_k {\bi v}_k$ and 
${\bi f}_k= -\p {\cal H}/\p {\bi r}_K$ with  $m_k$ being  
the mass of bound particle $k$. 
The coefficients  
 $\zeta_K(t)$ are fluctuating  friction constants  
 obeying  
\be 
\tau_{\rm h}^{2}{\dot \zeta}_K= {2\beta}E_K/dM     -1,
\en 
where $\tau_{\rm h}$ is the  characteristic time common to the two 
thermostats.  In this section, we fix 
$\beta=1/k_BT$. We assume that $\tau_{\rm h}$ 
is a short time independent of $T$ (see arguments around Eqs.(34) and (48)). 
 We write the kinetic energy of the bound particles in the layer $K$ as      
\be 
E_K= \sum_{k\in K}\frac{1}{2m_k } |{\bi p}_k|^2.
\en  
From Eq.(16) $E_K$ depends on $t$ as 
\be 
{\dot E}_K= 
\sum_{k\in K} {\bi v}_k\cdot{\bi f}_k  - 2\zeta_K E_K. 
\en

It is convenient to introduce    the  total   energy of  
the bound particles in the  layer K  as 
\be 
{\cal H}_K = E_K+ \sum_{k\in K} 
\Big[ \frac{1}{2}\sum_{j\in K}{\phi_{jk}}+
\sum_{i\le N}\phi_{ik}+\psi_{k}\Big].
\en   
The  total energy of the  particles in Eq.(2) is the sum, 
\be 
  {\cal H}= {\cal H}_{\rm in}+  {\cal H}_{\rm top}
+ {\cal H}_{\rm bot}.
\en  
From  Eqs.(11), (19),  and (20)
we  find  
\be 
{\dot{\cal H}}_K=  - {\cal I}_K 
 - 2\zeta_K { E}_K .
\en   
Therefore,  the  heat flow from the thermostat   $K$ 
to the particles (consisting of 
the bound ones  in the layer $K$ and the unbound 
ones) is written as   
\be 
{\cal J}_K = {\dot{\cal H}}_K + {\cal I}_K= - 2\zeta_K { E}_K. 
\en 
   From Eqs.(10) and (23)  $\cal H$ changes in time  as 
\be 
{\dot{\cal H}}=  {\cal J}_{\rm top}+{\cal J}_{\rm bot} . 
\en   
From  Eqs.(13), (23),  and (24) 
we now find  
\bea
&&\hspace{-5mm}
{\cal J}_{\rm top}= \frac{d}{dt}\Big[ 
\int d{\bi r}\frac{ z}{ H}{ \hQ} + {{\cal H}}_{\rm top}\Big] 
- \frac{1}{H}{\cal G}, \nonumber\\
&&\hspace{-5mm}  
{\cal J}_{\rm bot}=\frac{d}{dt}\Big [
 \int d{\bi r}(1-\frac{ z}{ H})
{\hQ} + {{\cal H}}_{\rm bot}\Big]
+ \frac{1}{H}{\cal G},
\ena 
which are our desired  surface-to-bulk relations.

{\subsection{ Liouville equation and equilibrium distribution}}

We denote all the    degrees of freedom in  our system by 
 the symbol $
 \Gamma= ({\bi r}_i,{\bi p}_i, \zeta_{K})
$  for the $ N+2M$ particles and the two thermostats. 
We also introduce  the sign reversal of 
the momenta and the friction constants, 
\be 
\Gamma= ({\bi r}_i, {\bi p}_i,\zeta_K) \to 
\Gamma^{*}= ({\bi r}_i, -{\bi p}_i, -\zeta_K).
\en 
In our system   the time-reversal symmetry holds. 
As a result, for each phase-space trajectory segment $\Gamma_t$ ($0<t<t_0)$, 
we can conceive its time-reversed conjugate 
 $\Gamma_t'= (\Gamma_{t_0-t})^*$, 
where $t_0$ is a fixed  time.

 The phase-space  
distribution $P(\Gamma, t)$   is  governed by  
   the Liouville equation ${\dot P}+ {\cal L} P=0$ 
in the Euler description in the phase space\cite{Mor}, where   
\be 
{\cal L}=\frac{\p }{\p \Gamma}\cdot{\dot \Gamma}= 
\sum_i \Big[\frac{\p}{\p {\bi r}_i}\cdot{\bi v}_i 
+  \frac{\p}{\p {\bi v}_i}\cdot{{\dot{\bi v}}_i}\Big]
+ \sum_{K}\frac{\p}{\p \zeta_K} {\dot\zeta}_K, 
\en 
For any initial distribution $P(\Gamma,0)$, we obtain  
\be
 P(\Gamma,t)= 
 \int d\Gamma_0 \delta (\Gamma -\Gamma_t) P(\Gamma_0,0), 
\en  
using   trajectories  $\Gamma_t$ 
starting from   $\Gamma_0$ at $t=0$.   

When the phase-space velocity 
$ {\dot \Gamma}(t)$ depends 
only on the instantaneous position 
$\Gamma_t$ autonomously in  stationary  environment, 
$\cal L$ is independent of $t$  
leading to  the convenient expression 
  $P(\Gamma,t)=\exp[{-{\cal L}t}] P(\Gamma,0) $.    
Also phase-space trajectories are  written as 
$\Gamma_t = \exp(-t{\cal L}^{\rm T})\Gamma$ with 
 the starting  point being  $\Gamma$,  
 where    ${\cal L}^{\rm T} =
-{\dot\Gamma}\p/\p\Gamma $ is   the transpose  operator 
 of $\cal L$. 
For any phase-space function ${\cal B}(\Gamma)$, 
we have ${\cal B}(\Gamma_t)= \exp(-t{\cal L}^{\rm T}){\cal B}(\Gamma)$.  
Hereafter, this $ {\cal B}(\Gamma_t)$  will    be written simply  
as ${\cal B}(t)$. The time derivative of $\cal B$ is written as 
$ \dot{\cal B}=  -{\cal L}^{\rm T}{\cal B}$.

For stationary and homogeneous $T$, 
we assume that the system 
tends to   the following equilibrium 
distribution, 
\be 
P_{\rm e}(\Gamma) =\frac{\gamma}{2\pi} 
\exp\Big[ {\beta F(T)- \beta{\cal H}(\Gamma)} - \sum_K \frac{\gamma}{2}
  \zeta_K^2 \Big], 
\en    
where $F(T)$ is the Helmholtz free energy for all the particles. The 
stationality condition   ${\cal L}P_{\rm e}=0$ holds   for\cite{Nose,Hoover}   
\be 
\gamma =dM\tau_{\rm h}^2. 
\en  
where the factor $dM$ appears because a single thermostat is 
attached to $M$ particle  in each boundary  layer. 
 Hereafter,  $\av{\cdots}_{\rm e} 
=\int d\Gamma(\cdots) P_{\rm e}(\Gamma)$ denotes 
the equilibrium  average. 
Then, $\av{\zeta_K \zeta_{K'}}_{\rm e}=\delta_{KK'}/\gamma$ 
and $\av{{\dot \zeta}_K {\dot \zeta}_{K'} }_{\rm e}=2\delta_{KK'}
/\gamma\tau_{\rm h}^2$.  The  stationality in equilibrium  gives 
\bea 
&&\hspace{-3mm}
\av{{\cal B}(t+t_0){\cal C}(t_0)}_{\rm e}=\av{{\cal B}(t){\cal C}(0)}_{\rm e}
= \av{{\cal B}(0){\cal C}(-t)}_{\rm e},\nonumber\\
&&\hspace{-3mm}
\frac{d}{dt}
\av{{\cal B}(t){\cal C}(0)}_{\rm e}
=\av{{\dot{\cal B}}(t){\cal C}(0)}_{\rm e}
= -\av{{\cal B}(t){\dot{\cal C}}(0)}_{\rm e}.
\ena 
Equal-time correlation functions 
$\av{{\cal B}(0){\cal C}(0)}_{\rm e}$ 
will  simply be written as 
$\av{{\cal B}{\cal C}}_{\rm e}$, which 
have no    time-dependence.

From Eqs.(23) and (29) we have 
$\av{{\cal J}_K }_{\rm e}=0$ and 
\be 
\av{{\cal J}_K{\cal J}_{K'}}_{\rm e} =
\delta_{KK'} (k_BT)^2 (dM+2)  /\tau_{\rm h}^2.   
\en 
For $K=K'$, the right hand side is  large for 
large $M /\tau_{\rm h}^2$. In contrast,   
  we find   $\av{({\cal I}_K)^2}_{\rm e}\propto 
 k_B Tn A $ from Eq.(11), which is different from Eq.(32).  
If ${\cal B}(\{{\bi r},{\bi p}\})$ is a variable  
 detemined by the particle positions and momenta 
and not by   $\zeta_K$, 
we have  $ \av{{\cal B} {\cal J}_K }_{\rm e}=0$. 
For example, we obtain 
\be 
  \av{{\hat q} {\cal J}_K }_{\rm e}
= \av{{\cal H} {\cal J}_K }_{\rm e}=\av{{\cal H}_{K'} {\cal J}_K }_{\rm e}
=\av{ J_h^\alpha  {\cal J}_K }_{\rm e}=0.
\en 
However, 
 the time-correlation functions   
$\av{{\cal B}(t) {\cal J}_K(0)}_{\rm e}$ 
are   nonvanishing for $t\gg \tau_{\rm h}$ (from  Eqs.(41) and (42)  below)  
and should be nearly equal to  $\av{{\cal B}(t) {\cal I}_K(0)}_{\rm e}$.

In  the Nos\'e-Hoover   thermostating, the time 
$\tau_{\rm h}$ in Eq.(17)   determines 
the  thermalization speed within  the boundary layers. 
We assume  that  $\tau_{\rm h}$  
is shorter than the typical microscopic  time $\tau_{\rm m}$ 
in the bulk.  Then, the  temperature fluctuations
of the thermostats  ($\cong 2E_K/{d M k_B}-T)$ 
 decay to zero quickly\cite{Bone}. We also assume 
a large spring constant ($= d^2\psi/dr^2$) 
to rigidly  pin  the bound particles on the wall\cite{Kawasaki,Shiba}. Then,  
the potential part of ${\cal H}_K$ in Eq.(20) should also 
relax rapidly (see  the second paragraph at the beginning of  Sec.II). 
     Under  these conditions,  Eqs.(19) and (23) indicate   
 \be 
\zeta_K 
\cong -{\cal I}_K/2E_K, 
\quad {\cal J}_K \cong {\cal I}_K.
\en    
This should hold even when the boundary temperatures $T_K(t)$ 
vary  in time.  Thus, slow time-evolution of $\zeta_K(t)$ 
arises from  that of ${\cal I}_K(t)$. 
 Such   thermostatted  layers  can 
serve as  {\it isothermal boundaries} in  the hydrodynamic description.
In their   simulation on shear flow,  
 Hoover {\it et al.}\cite{HooverPRL}   assumed 
  the friction of the form   $\zeta=  
\sum_{i} {\bi p}_i \cdot {\bi f}_i/\sum_i |{\bi p}_i|^2$ 
 for a single  thermostat coupled to 
all the particles 
as  ${\dot{\bi p}}_i= {\bi f}_i -\zeta {\bi p}_i$. 
In this method,  $\sum_i |{\bi p}_i|^2$ 
is exactly fixed  at a constant.

\section{Linear response  }

Next, starting with  the equilibrium  
distribution  $ P_{\rm e}(\Gamma) $ in Eq.(29) at  $t=0$, we slightly 
change  the top and bottom temperatures  in Eq.(17)  from $T$  to  
$T+\delta T_{K}(t)$    for $t>0$. 
We calculate linear response to $\delta T_K(t)$.

\subsection{Small bondary temperature changes} 

 We define    the mean temperature deviation 
$\delta T_{\rm m}$ and the mean temperature    gradient  $\cal T$  by  
\bea 
&&
\delta T_{\rm m}=  (\delta T_{\rm top}+\delta T_{\rm bot})/2, \\
&& {\cal T}= (\delta T_{\rm top}-\delta T_{\rm bot})/H. 
\ena 
which can depend on $t(>0)$. 
We   fix   the  thermostat time   $\tau_{\rm h}$ (and $\gamma$)  
(see discussions around Eq.(48)).

To linear order, ${\cal L}$ in Eq.(27) is changed by
\be 
{\delta {\cal L}}(t)= \frac{2}{\gamma } 
 \sum_K {\delta \beta_K(t)}   
E_{K} \frac{\p}{\p \zeta_K}.
\en 
where $\delta\beta_K(t)\cong  -\delta T_K(t)/k_BT^2$. 
The   deviation  $\delta P(t)= P(t)-P_{\rm e}$  obeys the linearized equation, 
\be 
\frac{\p}{\p t} 
{\delta {P}}(t) + {\cal L}{\delta P}(t) = -{\delta {\cal L}}(t)P_{\rm e} .
 \en 
From $\p P_{\rm e}/{\p \zeta_{K}}= -\gamma 
{\zeta_{K}}P_{\rm e}$,  we thus obtain    
\be
\delta P(\Gamma,t) =  -  \int_0^t ds 
\sum_K {\cal J}_K(-s) {\delta \beta_K(t-s) }P_{\rm e}(\Gamma), 
\en 
with the aid of $ e^{-s{\cal L}}{\cal J}_K P_{\rm e}
= {\cal J}_K(-s)  P_{\rm e}$ (see Eq.(31)). 
Here, we use  backward trajectories $\Gamma_{-s}$ ($0<s<t)$ 
with $\Gamma_0=\Gamma$ in ${\cal J}_K(-s)$. Note that 
 the variables ${\cal J}_K$  have already appeared in Eqs.(23)-(25). 
From   the time-reversal  symmetry in Eq.(26),    
we change  $\Gamma_{-s}$ to  $(\Gamma_{s})^{*}$  and 
  ${\cal J}_K(-s)$ to $  -{\cal J}_K(s)$ in Eq.(39). 
Then, we obtain    
\be
 \delta{P(\Gamma^{*}, t)}= 
  \int_0^t ds 
 \sum_K {\cal J}_K(s){ \delta \beta_K(t-s)}{P_{\rm e}(\Gamma)} , 
\en   
 using   forward trajectories $\Gamma_s$ ($0<s<t$) 
with $\Gamma_0=\Gamma$.

Now, for any phase-space variable ${\cal B}(\Gamma)$, we consider 
its deviation $\delta {\cal B}={\cal B}- \av{{\cal B}}_{\rm e}$.
 From Eq.(39),  its nonequilibrium  average   is written to  
linear order    as\cite{Kubo} 
\be 
\delta {\bar{\cal B}}(t) =    - 
\int_{0}^t  ds \sum_K \chi_{{\cal B}K} (s) {{\delta\beta }_{K}(t-s)}.   
\en 
We use the  time-correlation    functions, 
\be
 \chi_{{\cal B}K} (t)= \av{{\cal B}(t) {\cal J}_K(0)}_{\rm e}
= \av{{\cal B}(0) {\cal J}_K(-t)}_{\rm e}  , 
\en 
which   should be   nearly equal to $\av{{\cal B}(t) {\cal I}_K(0)}_{\rm e}$ 
for $t\gg \tau_{\rm h}$. For   ${\cal B}= 
{\cal H}$, Eq.(41) gives 
the average energy change $\delta{\bar{\cal H}}(t)$, 
where with the aid of  Eq.(24) we have     
\be 
\chi_{HK}(t) = \av{{\cal H}(t) {\cal J}_K(0)}_{\rm e}
=\int_0^t ds \sum_{K'}   \chi_{KK'}(s).
\en 
For the  heat flows ${\cal J}_K$, we define their time correlations,    
\be 
\chi_{K{K'}}(t) =  \av{{\cal J}_K (t) {\cal J}_{K'}(0)}_{\rm e},
\en 
which are nearly equal to  $ \av{{\cal I}_K (t) {\cal I}_{K'}(0)}_{\rm e}$ 
for $t\gg \tau_{\rm h}$. 

Let us  consider   the   average heat flux  
  ${\bar J}_h^z= -\lambda {\cal T}$ in the steady states with  constant  
$\delta T_K$ for small boundary resistance.   The coefficient 
 $\lambda$ is  the thermal conductivity, 
 \bea 
&&\hspace{-8mm} \lambda= \frac{H}{2k_BT^2}
\int_0^\infty\hspace{-1.5mm} 
 dt 
\av{{J}^z_h({\bi r},t)[{\cal J}_{\rm bot}(0)-{\cal J}_{\rm top}(0)]}_{\rm e}
\nonumber\\
&&\hspace{-7mm}= \frac{1}{Vk_BT^2}
\int_0^\infty\hspace{-1.5mm} 
 dt \int\hspace{-0.5mm} d{\bi r}\hspace{-0.5mm} \int d{\bi r}' 
\av{{J_h^z({\bi r},t) J_h^z({\bi r}',0)}}_{\rm e} .
\ena 
The  first line follows  from Eq.(41), where the integrand 
depends on $t$ and $z$ but its time integral is independent 
of $z$.  In the second line, 
we obtain Green's expression\cite{Hansen,Zwan,Onukibook,Green,Mor} 
 using  Eq.(25) and  the relation   
$\int_0^\infty dt  \av{J_h^z({\bi r},t) {\dot{\cal C}}(0)}_{\rm e}=0$
 for ${\cal C}={\hat{q}}$ and ${\cal H}_K$.  
The integrand in Green's expression  
is the flux correlation function depending on ${\bi r}-{\bi r}'$ and $t$ 
in the thermodynamic limit $V\to\infty$.

When  ${\cal B}= 
{\hat a}({\bi r},t)$ is  a    space-dependent  variable   such as 
$ \hat{q}({\bi r},t)$ and  ${J}^z_h({\bi r},t)$ in Eq.(41),     
 its nonequilibrium average  $\delta {\bar{a}}(z,t)$ 
and the  correlation functions   
$\chi_{{a} K}(z,t)= 
\av{{\hat{\hat a} }({\bi r}, t) {\cal J}_{K}(0)}_{\rm e} 
$ obey the hydrodynamics\cite{Landau-f,Kada}
 and depend on $z$ and $t$ in the   film geometry. 
Here, as  will be shown  in  Appendix A,  there are two 
  timescales\cite{Za,Onukibook,Ferrell,Miura},   
\be 
t_{\rm a}= H/c,\quad t_D= H^2/4D,
\en 
where    $c$ is the sound speed and  $D=\lambda/C_p$ 
is the thermal diffusion constant with  
 $C_p=nT(\p s/\p T)_p$  being   the isobaric specific heat per unit volume. 
For realistic $H$  we  have 
$t_D/t_{\rm a} \sim  Hc/D  \gg 1$. Then, as a causality 
 effect, there should be 
 no response in the interior  before  arrival of 
sounds emitted  from the boundaries. We thus predict   
\be 
\chi_{{a} K}(z,t)=0 \quad (0<ct<z~ {\rm and}~ H-z).     
\en  
Recently, similar   causality  was    confirmed  numerically 
 in a low-$T$ glass  
for propagation of transverse sounds\cite{Kawasaki}.

 We are assuming  that the thermostat time $\tau_{\rm h}$ in Eq.(17) 
is independent of $T$. The boundary heat flows ${\cal J}_K$ in Eq.(23) then  
appear in the response functions in Eq.(42). This was also assumed 
   in our simulation 
 on  glasses\cite{Shiba} and by   Lepri {\it et al.}\cite{Lepri} 
in their simulation of heat conduction in one dimension. 
More generally,  when  $\tau_{\rm h}$  depends on $T$, 
${\cal J}_K$ in Eqs.(39)-(42) is    replaced    by    
\be 
{\cal J}_K'= {\cal J}_K- dM k_BT^2 \Big(\frac{\p }{\p T}\tau_{\rm h}^2
\Big) \zeta_K{\dot \zeta}_K, 
\en 
where   the second correction  term ($\propto d\zeta_K^2/dt$) is  small 
 for  $t\gg\tau_{\rm h}$ from Eq.(34). 
It  disappears in the time integral 
$\int_0^\infty dt \av{{\cal B}(t) {\cal J}'_K(0)}_{\rm e}
$ for  $\av{{\delta{\cal B}}\zeta_K^2}_{\rm e}=0$.

\subsection{Dissipation function in linear regime} 

We can rewrite the linear response relation (41) as 
\be 
\delta{\bar{\cal B}}(t) = 
\av{{\cal B}(t) {\Omega}(t)}_{\rm e},
\en 
 where  $\Omega(t)$ is  given  by 
\be
 {\Omega}(t)= - \int_0^t ds \sum_K {\cal J}_K(s)  \delta \beta_K(s).  
\en 
We here calculate   this 
function in the linear regime with   details  in Appendix B.
It is  one example of  the Evans and Searles 
dissipation function  in their nonlinear theory\cite{Evans1,Searles1}. 
 In Sec.IVA, it  will be derived  in  the nonlinear regime.

From Eqs.(41) and (42), we notice that 
 the average ${\bar\Omega}(t)$ and the variance of $\Omega(t)$ are related as 
\be 
{\bar{{\Omega}}}(t)=\frac{1}{2}  \av{{\Omega}(t)^2}_{\rm e}  \quad (t>0).
\en 
Since  ${\Omega}(t)$ fluctuates, we can  define its distribution,    
\be 
f(\omega,t)=\int d\Gamma_0 P_{\rm e}(\Gamma_0)
\delta ({\Omega}(t)-\omega).
\en 
For ${\bar\Omega}(t)\gg 1$, $\Omega(t)$ arises from 
many microscopic events, so 
$f(\omega,t)$  asymptotically tends to be  Gaussian  as   
\be 
f(\omega,t)
\propto  \exp[\omega/2 - \omega^2/4{\bar{\Omega}}(t)]. 
\en 
  The magnitude  
of   ${\bar{\Omega}}(t)$ increases 
with increasing the system size 
and/or the  duration time $t$.

We can also obtain  a  bulk form of   ${\Omega}(t)$ from 
 Eq.(25). Using ${\cal G}(t)$ in Eq.(14), we  write its 
time derivative as 
\bea 
&&\hspace{-7mm}
{{\dot\Omega}}(t)=- \int d{\bi r}  {\delta{\bar \beta}(z,t)} 
\frac{\p}{\p t}{\hat q}({\bi r},t)- \sum_K \delta{\beta}_K(t)
 {\dot {\cal H}}_K(t) \nonumber\\
&&\hspace{5mm} 
-{\cal G} (t)[\beta_{\rm bot}(t)-
\beta_{\rm top}(t)]/H. 
\ena  
Here, we set $\delta{\bar \beta}(z,t) 
= \delta{\beta }_{\rm top}(t) z/H + 
 \delta{\beta }_{\rm bot} (t)(1-z/H)$, which is equal to 
$- \delta    {\bar T}(z,t) /k_BT^2$ 
to linear order with 
 \be  
\delta  {\bar T}(z,t) = \delta T_{\rm bot}(t)+ 
{\cal  T}(t) {z}.  
\en 
The second law of thermodynamics suggests  the following 
(not rigorously). In the first two terms in Eq.(54),  
 ${\hat q}({\bi r},t)$ and ${\cal H}_K(t)$ 
should tend to increase locally upon heating  (decrease upon cooling).
Here,  $\delta  {\bar T}(z,t) $    represents 
the  tempearture deviation correctly near the walls 
and approximately  far from them  even  in transient states.   
 In the third  term,  the integrated 
heat flux ${\cal G}(t)$ 
  should tend to be negative (positive) 
when  $\delta T_{\rm top}$ is larger (smaller) than 
$\delta T_{\rm bot}$.  
Therefore, it is very probable 
that  ${{\Omega}}(t)$ is  positive  
 for most initial states in the canonical ensemble, 
especially  for large systems and/or at long times.

After  the typical microscopic time $t_{\rm m}$, 
we can calculate   the nonequilibrium  average 
 ${\bar{{\Omega}}}(t)$  from 
the hydrodynamics (see Appendix B). 
Using the hydrodynamic deviation of the 
temperature   $\delta T({\bi r},t)$, 
that of  the density $\delta n({\bi r},t)$, 
 and  that of the velocity field ${\bi v}({\bi r},t)$, we 
can express ${\bar{\Omega}}(t)$    as 
\be 
{{\bar{\Omega}}(t)} =
\int \hspace{-1mm} d{\bi r}
 \Big[\frac{\rho |{\bi v}|^2}{2k_BT} - \frac{(\Delta S)_2}{k_B} 
+\int_0^t dt \frac{{\dot\epsilon}_{\rm th} +
{\dot\epsilon}_{\rm v}}{k_BT}\Big]>0.
\en
Here, ${\rho |{\bi v}|^2}/2$ is the hydrodynamic  kinetic energy 
 density with $\rho=mn$ and  
$(\Delta S)_2$ is the negative,   
second-order entropy deviation written as\cite{Callen,Landau-s,Onukibook}, 
\be 
(\Delta S)_2 = -\frac{ 1}{2T^2} C_V  (\delta T)^2 
- \frac{1}{2n^2T \kappa_T} (\delta n)^2<0, 
\en 
where    $C_V=(\p e/\p T)_n$ is  the  isochroic 
specific heat  per unit volume and $\kappa_T= (\p n/\p p)_T/n$ 
is the isothermal compressibility.  In the last term in Eq.(56), 
 $ {\dot\epsilon}_{\rm th}$ is the heat production rate 
per unit volume 
due to $\lambda$  and $ {\dot\epsilon}_{\rm v}$ is that  due 
to the viscosities\cite{Landau-f}: 
\bea 
&& {\dot\epsilon}_{\rm th}
= \lambda |\nabla \delta T|^2/T, \nonumber\\
&&\hspace{-1.2cm}
{\dot \epsilon}_{\rm v}
= \frac{\eta}{2}\sum_{ij} (\nabla_i v_j + \nabla_i v_i -\frac{2}{d}\delta_{ij} 
\nabla\cdot {\bi v})^2 
+ \eta_b| \nabla\cdot {\bi v}|^2 ,
\ena 
where   $\nabla_i v_j= 
\p v_j/\p x_i$ ($i,j=x,y, z$), $\eta$ is the shear viscosity, and $\eta_b$ is 
the bulk viscosity.

The physical meaning of $(\Delta S)_2$ is as follows. We 
 treat the thermodynamic entropy 
$S=ns$ per unit volume as a function of the energy density $e$ 
and the number density $n$. We then  superimpose  small deviations 
$\delta e$ and $\delta n$ on $e$ and $n$. 
Up to second order we  
 find\cite{Landau-s,Callen,Onukibook}  
\bea 
&& \hspace{-1.1cm}
(\Delta S)_2= 
S(e+\delta e, n+\delta n)- S(e,n)- (\delta e -\mu \delta n)/T \nonumber\\  
&&= [\delta e\delta (1/T) - 
 \delta n \delta(\mu/T)]/2,
\ena 
where $\mu$ is the reference   chemical potential and the second line 
 leads to Eq.(57). 
Here,    $\delta e$ and $\delta n$ 
are coarse-grained  variables varying smoothly 
in space. 
We can also  treat $\delta e$,  $\delta n$, and $\bi v$ as 
 local thermal fluctuations. For small deviations, their  distribution 
is given by\cite{Landau-s,Callen,Onukibook}     
\be 
\hspace{-0.4mm}
P_{\rm hyd}= {\cal N}  \exp\Big[ \int\hspace{-1mm}
 d{\bi r}\Big({\frac{(\Delta S)_2}{k_B} 
 -\frac{\rho |{\bi v}|^2}{2k_BT}}\Big)\Big],
\en 
where $\cal N$ is the normalization factor. 
The  {tempearture fluctuation} is defined by    
  $\delta T\equiv  C_V^{-1}[\delta e-(\p e/\p n)_T\delta n]$ 
 with 
     $\av{\delta T({\bi r})\delta T({\bi r}')}_{\rm e}= 
k_BT^2C_V^{-1} \delta({\bi r}-{\bi r}')$.  
From  Eqs.(56) and (60), ${\bar\Omega}(t)$ is rewritten as  
\be
 {{\bar{\Omega}}(t)} = - \ln( P_{\rm hyd}/{\cal N})+ \int_0^t dt
\int \hspace{-1mm} d{\bi r}\beta  ({{\dot\epsilon}_{\rm th} +
{\dot\epsilon}_{\rm v}}).
\en 
In  Eqs.(56), (57), and (61)    
time-dependence of $\delta T_K(t)$ 
can be arbitrary.

To explicitly calculate ${\bar\Omega}(t)$,  we assume  $  
\delta T_K(t)= \theta (t) \delta T_K,
$  where  
 $\theta(t)$ is  the Heaviside  step function. 
In the initial satage  $0<t<t_{\rm a}$, 
the disturbances are localized near the walls and  
${\bar{\Omega}}(t)$ grows algebraically   
and ${\bar\Omega}(t)\propto A$ (see  Eq.(B3)).   
For $t> t_{\rm a}$,  we 
calculate   ${\bar\Omega}(t)$ as  
 \be 
\frac{
{\bar{{\Omega}}}(t)}{V} 
=\psi_1(\tau) C_V \frac{ {(\delta T_{\rm m})^2}}{k_BT^2}  
 + \Big(t+ t_D\psi_2(\tau) \Big) \frac{\lambda {\cal T}^2 }{k_BT^2}, 
\en 
where   ${\bar{{\Omega}}}(t) \propto  V=AH$.  The    
$\psi_1(\tau)$ and $\psi_2(\tau)$ are monotonically increasing,  positive 
functions of the scaled time $\tau= t/t_D= 4D t/H^2$ 
with     $\psi_1(0)= \psi_2(0)=0$, $\psi_1(\infty)=1$, 
and $ \psi_2(\infty)=1/3$. As a unique effect,  
  $\psi_1(\tau) $ tends  to 1 quickly on the 
 {\it piston time} 
($\sim t_D/\gamma_{\rm s}^2$) for   $\gamma_{\rm s}=C_p/C_V\gg 1$ 
near the criticality\cite{Za,Onukibook,Ferrell,Miura}. 
   At long times ($>t_D$), we have ${\bar{\Omega}}(t)  \cong  
(\epsilon_{\rm th}/k_BT)Vt $.

In  Kubo's  theory\cite{Kubo},  
 the disssipation function  is given by  
$\Omega(t)=\beta \int_0^t ds {\dot{\cal A}}(s) \gamma_{\rm ex} (s)
= \beta({\cal H}(t)-{\cal H}(0))$ 
for the total Hamiltonian     
${\cal H}'= {\cal H}- \gamma_{\rm ex} (t) {\cal A}$ 
without  thermostats, where $\gamma_{\rm ex} (t)$ is applied  for $t>0$. 
Then,  Eq.(51)  holds in the linear regime, 
while Eqs.(94)-(96) hold  in the nonlinear regime 
(if  the perturbation ${\cal H}'-{\cal H}$ remains  unchanged).  
Thus,  we generally  have 
${\bar\Omega}(t)=\beta \delta{\bar{\cal H}}(t)>0$.

\subsection{Time-correlations with ${\cal J}_K$ }

In  the early hydrodynamic  stage $t_{\rm m}\ll  t<t_{\rm a}$, 
 Eq.(A4) and (A5) indicate that 
 the time-correlation functions  ${ \chi_{HK}(t)}$ and ${ \chi_{KK}(t)}$ in Eqs.(43) and (44) behave as 
\be 
{ \chi_{HK}(t)}\cong   \int_0^t ds 
{ \chi_{K{K}}(s)}\sim {k_BT^2A} {\lambda}/{\sqrt{Dt}}. 
\en  
which increases as $t^{-1/2}$ as  $t \downarrow t_{\rm m}$. 
  Thus,    $ \chi_{K{K}}(t)$ 
assume large positive values for $0<t\ls t_{\rm m}$ as in Eq.(32), 
but they are negative for $t\gg t_{\rm m}$. 
For $K\neq K'$,  we have $ \chi_{K{K'}}(t)=0$  in 
the time range $0<t<t_{\rm a}$ for $H\gg D/c$ from Eq.(47). 
For $t> t_{\rm a}$, Eq.(A13) leads to   
\be 
\int_0^t ds 
\frac{ \chi_{K{K'}}(s)}{Ak_BT^2} 
 = 
 \frac{\lambda\psi_1'}{H\gamma_{\rm s}} +(2\delta_{K{K'}}-1)
 (1+\psi_2')\frac{\lambda}{H} ,
\en 
where   $\psi_1'= d\psi_1(\tau)/d\tau$ and 
$\psi_2'= d\psi_2(\tau)/d\tau$ (see Appendices A and  B). 
The first term in Eq.(64) 
behaves as $\lambda (1-\gamma_{\rm s}^{-1}) /\sqrt{4\pi Dt}$ 
for $t_{\rm a}<t<t_D/(\gamma_{\rm s}-1)^2$ from Eq.(B7) 
and is continuously conected to Eq.(63).

In the limit $t\to\infty$, Eq.(64)  becomes
\be 
\int_0^\infty dt {\chi_{K{K'}}(t) }  
=(2\delta_{KK'} -1)k_BT^2\lambda {A}/{H} . 
\en 
In terms of   ${\cal J}_{\rm a}=({\cal J}_{\rm bot}- 
{\cal J}_{\rm top})/2 $, we 
obtain    the surface expression for $\lambda$ 
derived  by Petravic and Harrowell\cite{Harrowell},   
\be 
\lambda= \frac{H}{k_BT^2 A} 
 \int_0^\infty dt \av{{\cal J}_{\rm a}(t) 
{\cal J}_{\rm a}(0)}_{\rm e}. 
\en  
In Eqs.(63)-(66)  we can  replace   
$\av{{\cal J}_K (t) {\cal J}_{K'}(0)}_{\rm e}$ 
by  $\av{{\cal I}_K (t) {\cal I}_{K'}(0)}_{\rm e}$ from Eq.(34).  
These integral relations stem from 
 the hydrodynamics, while  the short-time behavior 
of  $ \chi_{K{K}}(t)$  depends on the thermostating method. 
In Eqs.(65) and (66) (and the first line of Eq.(45)), 
 the main  contributions  
arise from the  long time range  $t\sim t_D$.

To explain the thermal resistance of   a solid-liquid   interface 
 in $^{3}$He,   Puech {\it et al.}\cite{Cast} 
expressed the Kapitza length $\ell_{\rm K}$ as   
$\lambda/\ell_{\rm K} = \int_0^\infty dt 
\av{{\cal J} (t) {\cal J}(0)}_{\rm e}
/ (k_BT^2 A)$ using the Onsager theory\cite{Onsager,Landau-s}, 
where  ${\cal J}(t)$ is the microscopic  heat flow  
through the interface. 
Barrat   and   Chiaruttini\cite{Barrat} 
calculated  the integral $G(t)= \int_0^t  ds 
\av{{\cal J} (s) {\cal J}(0)}_{\rm e}/k_BT^2A$ 
 at a surface between a solid 
and a Lennard-Jones liquid, which assumed a plateau 
after  a microscopic time  
and decayed  slowly. 
They identified the plateau  as 
$\lambda/\ell_{\rm K}$, 
while  Eq.(65) indicates  
$\lim_{t \to \infty}G(t) = \lambda/H$ for finite $H$.   
This aspect  should further be investigated 
in future.

\subsection{ Steady-state relations  using surface heat flows}

In the steady state at  constants $\delta T_K$,  the expression for 
the  average follows from  Eqs.(23) and (41) as 
\be 
\delta {\bar{\cal B}} = - \sum_K  
\av{\delta B({ U}_{K\infty}+\delta{\cal H}_K )}_{\rm e}
 {{\delta \beta}_{K}}.
\en 
Using backward trajectories $\Gamma_{-s}$ 
with $\Gamma_0=\Gamma$, we define  
\be 
U_{K\infty}= \int_0^{\infty}  ds e^{-\epsilon s}  {\cal I}_K(-s),
\en 
where $\epsilon$ is a positive small  number 
ensuring convergence of the time integral. 
Up to linear order, the steady-state distribution  
$ P_{\rm st}(\Gamma)$ is expressed as  
\be 
 P_{\rm st}(\Gamma) /P_{\rm e}(\Gamma)
=1 - \sum_K ({ U_{K\infty}+ \delta 
{\cal H}_{K}})\delta \beta_K  . 
\en  
In particular, if  $\delta\beta_{\rm top}=\delta\beta_{\rm bot}
=- \delta T/k_BT^2$, 
$P_{\rm st}(\Gamma)$ is the new equlibrium distribution with 
the shifted temperature $T+\delta T$, since  
 Eq.(10)  gives    
$
\sum_K U_{K\infty}= 
{\cal H}_{\rm in}- \av{{\cal H}_{\rm in}}_{\rm e}.
$  
From  the time reveral symmetry,   we also obtain  
\be 
 P_{\rm st}(\Gamma^*)/P_{\rm e}(\Gamma)= 1
- \sum_K ({ U'_{K\infty}+ \delta 
{\cal H}_{K}})\delta \beta_K  ,
\en 
where forward trajectories  appear as in Eq.(40) with  
\be  
 U'_{K\infty}
= - \int_0^{\infty}  ds e^{-\epsilon s}  {\cal I}_K(s) .
\en

We also consider 
 the steady-state averages ${\bar{\cal J}}_K$,  
 ${\bar{\cal I}}_K$, and 
 ${\bar J}_h^z$.  From  Eq.(5) we find     
\be 
\frac{d}{dz} {\bar J}^z_h(z) = 
\AV{ \sum_{i\le N} 
{\bi v}_i\cdot ({\bi f}_i^{{\rm top}}+{\bi f}_i^{{\rm bot}})
   \delta(z- z_i)}_{\rm s} \frac{1}{A} , 
\en 
where we set $\delta({\bi r}-{\bi r}_i)\to 
\delta({z}-{z}_i)/A$ and $\av{\cdots}_{\rm s}$ denotes 
the steady-state average.   
Thus,    ${\bar J}_h^z $   
is a constant in the interior   
but depends on $z$ near the boundaries 
decaying  to zero in the walls. Integrating   
Eq.(72) across the boundaries and using Eqs.(13) and (25), we obtain   
\be 
 {\bar{\cal J}}_{\rm top}=
{\bar{\cal I}}_{\rm top}=
- A{\bar J}_h^z, \quad 
 {\bar{\cal J}}_{\rm bot}={\bar{\cal I}}_{\rm bot}=
 A{\bar J}_h^z.
\en 
where ${\bar J}_h^z= -\lambda 
{\cal T}$ for  negligible  boundary resistance.

\subsection{ Steady-state bulk relations }

Using   Eq.(25), 
we obtain the bulk expression 
for the steady-state average, 
\be
 {{ \delta {\bar{\cal B}}} } 
=  \int d{\bi r}' \gamma_B({\bi r}'){\delta {\bar T} (z')} 
 +\sum_K    \gamma_{BK}{\delta T_K} 
-\chi_B^h{  \cal T}.   
\en 
where ${\delta {\bar T} (z)}$ is the temperature profile in Eq.(54). 
The first two   terms are local-equilibrium parts     
written in terms of  equal-time correlations as   
\bea 
&&\gamma_B({\bi r}) =
\av{{\delta{\cal B}}\delta\hQ({\bi r})}_{\rm e}/{k_BT^2},\\
&&  \gamma_{BK} = \av{{\delta{\cal B}}\delta{\cal H}_K}_{\rm e}/{k_BT^2}. 
\ena 
which  vanish if $\cal B$ is odd with respect to the time reversal.

The third   term in Eq.(74) is  dissipative. 
In terms of  ${\cal G}(t)$ in Eq.(14),  
  $\chi_B^h$ is given by    
\be 
\chi_B^h= \int_0^\infty  dt 
{\av{\delta{\cal B}(t) {\cal G}(0)}_{\rm e}}/k_BT^2,
\en  
which yields  Green's expression (45)  for     ${\cal B}= {\cal G}$. 
If ${\cal B}(t)$ is a  long-wavelength 
hydrodynamic variable, it evolves 
 slowly in time   remaining nearly  orthogonal to 
$J_h^x(0)$. For this case, 
we can  neglect  the  dissipative term in Eq.(74) 
at long wavelengths\cite{Mori,Zwanzig1961}.

We  further examine    the first  term in Eq.(74)    
when  ${\cal B}= {\hat a}({\bi r})$ is   
a space-dependent variable having the 
 even  time-reversal symmetry. In this case, we 
should replace $\gamma_B({\bi r}')$  by   
 the   two-point equal-time correlation function,   
\be 
\gamma_a({\bi r}, {\bi r}')= 
{\av{\delta {\hat a}({\bi r}) \delta {\hQ}({\bi r}')}_{\rm e}}
/{ k_BT^2},
\en 
while  the second 
term ($\propto \av{{\delta{\hat a}({\bi r})}\delta{\cal H}_K}_{\rm e}$)
in Eq.(74) vanishes  far from the boundaries.  
We treat the equilibrium average  
$a\equiv  \av{{\hat a}}_{\rm e}$ 
as a thermodynamic quantity depending $T$ and $p$. 
In Appendix C, we will find the  behavior, 
\be  
\gamma_a({\bi r}, {\bi r}')= 
g_{a} ({\bi r}- {\bi r}') +  \ppp{a}{p}{T}\ppp{p}{T}{n}\frac{{ 1}}{V}, 
\en   
where  ${\bi r}$ and ${\bi r}'$ are  far from the 
boundaries. The function  
$g_{a} ({\bi r})$ is  short-ranged    satisfying  
\be 
\int d{\bi r}g_{ a}({\bi r})=\ppp{a}{T}{p}. 
\en 
The   second term in Eq.(79)  is  proportional to  $V^{-1}$. Thus,   
\be 
\int d{\bi r}\gamma_{ a}({\bi r},{\bi r}')=\ppp{a}{T}{n}.
\en 
We assume that the correlation length 
of $g_{ a}({\bi r})$ is much shorter than $H$. 
The  long-range behavior 
 ($ \propto V^{-1})$ in the density  correlation functions   has  
 been studied  in the canonical 
ensemble\cite{Rogers,Kremer,Vlugt,Lebo}.
In Appendix C, we will examine it for  general 
correlation functions 
${\av{\delta {\hat a}({\bi r}) \delta {\hat b}({\bi r}')}_{\rm e}}$.

We now substitute Eq.(79) into the first term in 
Eq.(74).  Using  Eq.(80) we obtain the 
local-equilibrium part of the 
 steady-state average  far from the boundaries, 
\be
 {{ \delta {\bar{ a}}} }_{\rm loc}(z) 
=   \ppp{a}{ T}{p} \delta  {\bar T}(z)  
+ \ppp{a}{ p}{T} \delta {\bar p} ,
\en 
where $\delta {\bar T}(z')$ is replaced by $\delta {\bar T}(z)$ 
in   $\int d{\bi r}' g_a({\bi r}- {\bi r}'){\delta {\bar T} (z')} $ 
and $\delta {\bar p}$ is  the homogeneous pressure deviation,   
\be 
\delta {\bar p}=({\p p}/{\p T})_{n}  \delta T_{\rm m}. 
\en 
If  $\hat a$ is equal to $  \hat{p}$ in Eq.(8), 
  its  average deviation 
is equal to  the above $\delta {\bar p}$.  
For  ${\hat a}={\hat n}$ and ${\hat q}$, 
 we  obtain  
\bea
&&\delta {\bar n} (z)= (\p n/\p T)_p   (\delta  {\bar T}(z)- \delta T_{\rm m} ), \\
&&nT \delta {\bar s}(z)= C_p (\delta  {\bar T}(z)- \delta T_{\rm m} )
+  C_V \delta T_{\rm m} ,
\ena 
Thus,   the space average of 
$\delta {\bar n} $ is zero and that of $nT \delta {\bar s}$ 
is  $ C_V \delta T_{\rm m}$.  
We confirm that     $\delta  {\bar T}(z)$ introduced in Eq.(55) 
is    the local  temperature deviation   in steady states.

As another application of Eq.(77), 
we can  set ${\cal B}(t)= {\hat a}({\bi r},t) 
 {\hat b}({\bi r}',t) $, where $\hat a$ and $\hat b$ 
are hydrodynamic variables. Then, we  obtain 
the steady-state pair correlation 
$g_{ab}^{\rm s}({\bi r}, {\bi r}')= \av{ {\hat a}({\bi r},t) 
 {\hat b}({\bi r}',t)}_{\rm s}$. To linear order in $\cal T$, 
it depends only on 
$r= |{\bi r}- {\bi r}'|$ far from the boundaries and  its   deviation 
 is  long-ranged as $ {\cal T} r^{-1} $ 
for $d=3$ and as $ {\cal T} \ln r $ 
for $d=2$ due to the mode-coupling 
effect\cite{Onukibook,mu,Oppen,Dorf,Gunton}.

\section{Nonlinear theory} 
  
We finally  study   nonlinear dynamics, where 
 $\delta T_K(t)$ are time-dependent and need  not    small.

\subsection{Phase-space distribution}

The trajectory equation is   non-autonomaous    as  
\be 
{\dot \Gamma}_t= {\cal V}(\Gamma_t, t),
\en  
where  ${\cal V}(\Gamma, t)$ is 
the phase-space velocity in the Euler description. 
It depends on $t$ in addition to $\Gamma$, since 
the equations for $\zeta_K(t)$ are 
changed from Eq.(17) to 
\be 
\tau_{\rm h}^{2}{\dot \zeta}_K(t) 
=2 \beta_K(t) E_K(t) /dM      -1, 
\en  
where  $ \beta_K(t) = 1/k_B[T+ \delta T_K(t)]$ and $\tau_{\rm h}$ 
is a constant.

 The phase-space  distribution $P(\Gamma, t)$ 
obeys the Liouville equation and 
is still expressed  in  the form of  Eq.(28) 
in terms of the initial distribution $P(\Gamma,0)$. Using   trajectories  
 starting  from $\Gamma_0$  and reaching   
$\Gamma_t$, we  find   
\be
 P(\Gamma_t,t)=  e^{-{\Xi}(t)} P (\Gamma_0,0). 
\en 
Here,  $e^{{\Xi}( t)}$  is equal to 
the Jacobian $ d\Gamma_t/d\Gamma_0$, where  
 the phase-space volume element moves 
from $d{\Gamma}_0$  to  $d{\Gamma}_t $.    
We can then rewrite  Eq.(88) as 
\be 
d\Gamma_t P(\Gamma_t, t)=d\Gamma_0 P(\Gamma_0, 0). 
\en 
Generally, ${\dot{\Xi}}(t)$ is  equal to the time integral of 
 the phase-space expansion   factor $\Lambda(t)$  as 
\be 
{\Xi}(t)= \int_0^t ds  \Lambda(s).
\en
where $\Lambda(t)=  \Lambda(\Gamma_t,t)$ 
with  $\Lambda(\Gamma,t)
= (\p {\cal V}/\p{\Gamma})_t$. 
 For  the  Nos\'e-Hoover thermostating, Eq.(16) gives        
 \be
{\Lambda}(t)=-dM   \sum_K\zeta_K(t).
\en

For any initial distribution $P(\Gamma,0)$, Evans and Searles 
defined  the dissipation function $\Omega(t)$ by\cite{Evans1,Searles1} 
\be 
\ln [P(\Gamma_0,0)/P(\Gamma_t ,0)]= \Xi(t)+\Omega(t).
\en 
Let   $P(\Gamma,0)$ be 
equal to the equilibrium distribution   $P_{\rm e}(\Gamma)$ 
in Eq.(29). Then, time derivative of  Eq.(92) gives  
\bea   
{\dot\Omega}(t)&&= 
\frac{d}{dt}
\Big [ \beta{\cal H}(t)
+ \sum_K \frac{\gamma}{2}\zeta_K(t)^2\Big]- 
{\Lambda }(t)\nonumber\\
&&
= - \sum_K \delta\beta_K (t) {\cal J}_K(t),
\ena 
where ${\cal H}(t)={\cal H}(\Gamma_t)$.  This   agrees  with Eq.(50). We  
find 
\be 
d\Gamma_t
P_{\rm e}(\Gamma_t) 
=d\Gamma_0 P_{\rm e}(\Gamma_0) \exp[-{\Omega}(t)].
\en 
Phase-space integration of Eq.(94) gives 
\be 
\av{e^{-\Omega(t)}}_{\rm e}=\int d{\Gamma}_0 P_{\rm e} (\Gamma_0)
e^{-\Omega(t)}= 1.
\en 
Here, if  $e^{-\Omega(t)}$ is expanded 
 with respect to  $\Omega(t)$, we obtain    
 $\av{\Omega(t)}_{\rm e}= \av{\Omega(t)^2}_{\rm e}/2+\cdots$, 
which coincides with  Eq.(51) in second order. We also  
find ${\bar \Omega}(t)>0$ since $e^{-x}+x\ge 1$ holds 
for any $x$.  From  Eqs.(89) and (94) we also obtain  
\be
P(\Gamma,t)=  P_{\rm e}(\Gamma)  \exp[{\Omega}(t) ],
\en 
where the  trajectories  $\Gamma_s$ ($0<s<t$) in $\Omega(t)$ 
end  at $\Gamma_t= \Gamma$ with the initial points $\Gamma_0$ 
being a function of $\Gamma$. Note that     $\Omega(t)$ 
satisfies    Eqs.(94)-(96) for any  $P(\Gamma,0)$. 
 See  another  choice of  $P(\Gamma,0)$ in Sec.IVE.

Evans and Searles\cite{Evans1,Searles1,W}   obtained a different form 
of $\Omega(t)$, 
where ${\cal J}_K(t) =-2E_K(t)\zeta_K(t)$ 
in our $\Omega(t)$  is replaced by 
$- dM k_BT_K(t)   \zeta_K(t) $. 
Thus, there is no essential 
difference between   our  $\Omega(t)$  and theirs  
on long timescales($\gg \tau_h$)  from Eq.(34). 
In our scheme, their result exactly  follows for   
$\tau_{\rm h}^2 \propto T^{-1}$. 
In  Appendix D,  we will derive $\Omega(t)$ 
 generally  including $T$-dependence of  $\tau_{\rm h}$.

\subsection{Transition between  equilibrium states}

Let the two boundary walls have the same 
temperature $T(t)= T+\delta T(t)$  with $\delta T(0)=0$, 
which tends  to the final one $T_{\rm f}=T+\delta T_{\rm f}$ 
for $t\gg t_{\rm ex}$. We set $\delta\beta(t)\equiv   \delta\beta_{\rm top}(t)
=  \delta\beta_{\rm bot}(t) $ and use  Eq.(24) to  find the simple form, 
\be 
{\Omega}(t)= - \int_0^t ds 
{\dot{\cal H}}(s)  \delta\beta(s).  
\en  

It is convenient to  define the following  {\it excess} function,   
\be 
{\Omega}_{\rm ex}(t)
=  {\Omega}(t)+  {\cal H}(t)  \delta\beta(t)=  
 \int_0^t ds  {\cal H} (s) {\dot\beta}(s), 
\en 
where    ${\dot\beta}(t)=d{\beta}(t)/dt$ and ${\Omega}_{\rm ex}= 
{\cal H}(0)\delta\beta$  for the stepwise change.
If ${\dot\beta}(t)= 0$ for $t>t_{\rm ex}$, 
${\Omega}_{\rm ex}(t)$ is independent of $t$ 
for $t>t_{\rm ex}$ for each  $\Gamma_0$.  We  rewrite  Eq.(94)   as 
\be 
d\Gamma_t
{ P_{\rm e}(\Gamma_t)}e^{ - {\cal H}(t)  \delta\beta(t)}
=d\Gamma_0  P_{\rm e}(\Gamma_0) e^{ 
-{\Omega}_{\rm ex}(t)}.  
\en 
The  left hand side is proportional to the canonical 
distribution at 
the temperature $T(t)$, which is  expressed    as 
\be 
P_{\rm e}(\Gamma; T(t)) = { P_{\rm e}(\Gamma)}\exp[{{\cal F}(t) 
- {\cal H}(\Gamma)  \delta\beta(t)}].
\en 
Here, the factor $\exp[{\cal F}(t)]$ arises 
from the normalization condition. 
Phase-space integration of Eq.(99) gives  
\bea 
&&\hspace{-18mm}{\cal F}(t)=  
 \beta (t) F(T(t))-\beta F(T) \nonumber\\
&&\hspace{-1cm} 
= -\ln \Big[  \av{e^{ 
-{\Omega}_{\rm ex}(t)}}_{\rm e} \Big],
\ena 
where $F(T)$ is the Helmholtz free energy for all the particles. 
More generally, for any variable ${\cal B}(\Gamma)$, we consider 
its {\it equilibrium average} 
at the temperature $T(t) $, written as 
$b(T(t))=  \int d\Gamma  {\cal B}(\Gamma) P_{\rm e}(\Gamma;T(t))$, 
which changes from $b(T)$ at the initial temperature $T$ 
to $b(T_{\rm f})$  at the final one $T_{\rm f}$  as $t$ increases. 
We multiply   Eq.(99) by ${\cal B}(t)= {\cal B}(\Gamma_t)$  and 
 perform  its  phase-space integration to  obtain  
\bea
&&\hspace{-13mm}
b(T(t))  
=  e^{ {\cal F}(t)}
\av{{\cal B}(t)  e^{ 
-{\Omega}_{\rm ex}(t)}}_{\rm e} \nonumber\\
&&\hspace{-1mm} = b(T)+  e^{ {\cal F}(t)}
\av{\delta{\cal B}(t)  e^{ -{\Omega}_{\rm ex}(t)}}_{\rm e},
\ena 
where $\delta{\cal B}(t)={\cal B}(t)-b(T)$. 
 In  Eqs.(101) and (102),  
  the average is over $P_{\rm e}(\Gamma_0)$ 
with trajctories starting from 
 $\Gamma_0$.  We can  confirm Eq.(102) to linear order using 
 $\av{\delta{\cal B}(t)}_{\rm e}\cong  
\av{\delta{\cal B}(t) \Omega(t)}_{\rm e}
= \av{\delta{\cal B}(t)\Omega_{\rm ex}(t)}_{\rm e}-
\av{\delta{\cal B}{\cal H}}_{\rm e}\delta\beta(t)$, 
 $\av{\delta{\cal B}{\cal H}}_{\rm e}
= k_BT^2 (\p b/\p T)_n $, and $e^{-\Omega_{\rm ex}}\cong 1-\Omega_{\rm ex}$.

Previously, Williams {\it et al.}\cite{W} 
  obtained some general  relations and one of 
them  is equivalent to Eq.(101). 
We note that Eq.(101)  resembles 
  Jarzynski's equality\cite{Jar} for 
 isothermal transitions between two 
equilibrium states.

\subsection{Time reversal  }

For each  trajectory segment $\Gamma_s$ ($0<s<t)$, 
we can concieve  its time-reversed  conjugate: $\Gamma_s'$ 
with $\Gamma_0'= ( \Gamma_t)^* $ 
and $\Gamma_t'= ( \Gamma_0)^*$ at fixed  $t$ (see Eq.(26)).
In  the present non-stationary (non-autonomous) situation, the time-reversed 
friction  variables, written as  $\zeta'_K(s)$,  obey   
\be 
\tau_{\rm h}^{2}\frac{d}{ds}{ \zeta}'_K(s) 
=2 \beta_K(t-s) E'_K(s) /dM      -1.  
\en 
Here,   $\beta_K(t-s)$ appear. Then,   the time-reversed 
 dissipation function $\Omega'(t) $  is just equal to  $-\Omega(t) $, 
where  $\Omega(t) $  is   the original one. Assuming the equilibrium 
 distribution $P_{\rm e}(\Gamma'_0) (= P_{\rm e}(\Gamma_t))$ 
for the initial points $\Gamma'_0$, we define  
\be 
f_r(\omega,t) =\int d\Gamma_0' P_{\rm e}(\Gamma_0')
\delta (\Omega'(t)-\omega).
\en 
Here,    $d\Gamma_0' P_{\rm e}(\Gamma_0')= 
d\Gamma_t P_{\rm e}(\Gamma_t)= d\Gamma_0 P_{\rm e}(\Gamma_0)
e^{-\Omega(t)}$ from Eq.(94) 
and $\delta (\Omega'(t)-\omega)= \delta (\Omega(t)+\omega)$ from  
  $\Omega'(t)=-\Omega(t) $. Then, $f_r(\omega,t)$ is related to  
$f(\omega,t) $  in Eq.(52) as 
\be 
 f_r(\omega,t) = e^{\omega} f(-\omega,t) ,
\en 
which is  a well-known result\cite{Ko,Jar1,Cr}.

Furthermore, we find 
$f_r(\omega,t) =  f(\omega,t)$  when 
\be 
\delta\beta_K(s)=\delta\beta_K(t-s) \quad (0<s<t). 
\en 
This holds for stepwise changes. Under Eq.(106), we have 
\be 
 f(\omega,t) = e^{\omega} f(-\omega,t).
\en   
This is the transient fluctuation theorem by 
 Evans and Searles\cite{Evans1,Searles1,FD}, 
which  is exact  for   stepwise  changes  
of external parameters. It was  
checked in simulations of  small systems\cite{Evans1,Searles1,W}.  
We can also realize Eq.(106) for periodic $\delta T_K(t)$, 
  where Eq.(107) holds  for particular $t$. 
In the nonlinear regime, $f(\omega,t)$ can 
significantly deviate from 
  the Gaussian form in Eq.(53) at small $t$ 
due to  {events}  with large 
$|\Omega(t)|$\cite{Ri,Ha,Se}. 
When the Gaussian form is nearly realized with ${\bar\Omega}(t)\gg 1$, we have 
$e^{-\omega}f(\omega,t)\propto \exp[-(\omega+ {\bar\Omega})^2/4{\bar\Omega}]$, 
so  the equality (95) holds due to  {\it rare events}  with largely negative 
$\Omega(t)( \cong -{\bar{\Omega}}(t)$).

Furthermore, under Eq.(106),  Eqs.(28) and (95)  give  
\be 
P(\Gamma^{*},t)= P_{\rm e}(\Gamma ) e^{-\Omega(t)},  
\en 
where  we use forward  trajectories $\Gamma'_s$ 
 starting  from $\Gamma_0'= 
\Gamma$ in $\Omega(t)$. For example, from Eq.(108), 
the average of the heat flow  $J_h^z({\bi r},t)$ in Eq.(7) is   
written as 
\be
{\bar J}^z_h(z,t) 
=- \av{ J_h^z({\bi r},0)e^{-{\Omega}(t)}  }_{\rm e}.  
\en 
which  obeys the hydrodynamics and 
  satisfies   Eq.(47).

\subsection{Local-equilibrium distribution } 

For the stepwise boundary-temperature change, 
we have obtained Eqs.(107)-(109). 
We can further perform time-integration of  Eq.(54). 
Some calculations give 
\be 
\Omega(t)=\Psi(\Gamma_0)-\Psi(\Gamma_t) +
 {\cal D}(t) .
\en 
Using ${\cal G}(t)$ in Eq.(14) we introduce 
\bea 
&& \hspace{-7mm}
\Psi(\Gamma)= 
 \int d{\bi r}  {\delta{\bar \beta}(z)}  
{\hat q}({\bi r};\Gamma) 
+  \sum_K \delta{\beta}_K {\cal H}_K(\Gamma),\\
&&   {\cal D}(t)=  -\gamma_{\rm a} \int_0^t \hspace{-1mm}ds {\cal G}(s).
\ena 
Here, we write 
${\hat q}({\bi r};\Gamma_t)={\hat q}({\bi r},t) $   
to avoid confusion and  the coefficients  
 ${\delta{\bar \beta}(z)} $ and $\gamma_{\rm a}$ are   defined  by 
\be
{\delta{\bar \beta}(z)} =\delta\beta_{\rm bot}- \gamma_{\rm a}z,\quad 
 \gamma_{\rm a}= (\beta_{\rm bot}-
\beta_{\rm top})/H, 
\en  
where ${\delta{\bar \beta}(z)} \cong -{\delta{\bar T}(z)}/k_BT^2$ 
 and $\gamma_{\rm a} \cong {\cal T}/k_BT^2$  in linear order 
with ${\delta{\bar T}(z)}$ being given by  Eq.(55).  The ${\cal D}(t)$ 
represents the entropy production in the bulk region.  

Using $\Psi(\Gamma)$ in  Eq.(111) 
we can define the local-equilibrium distribution 
at fixed $N$ and $V$ in the form, 
\be
P_{\rm lc}(\Gamma)= C_{\rm lc} 
P_{\rm e} (\Gamma) \exp[- \Psi(\Gamma)], 
\en 
where the local  inverse temperature is given 
by $\beta+\delta{\bar \beta}(z)$ and  
 $C_{\rm lc}$ is the  normalization constant determined  by  
\be 
C_{\rm lc}= \av{e^{-\Psi(\Gamma)}}_{\rm e}^{-1} 
= \av{e^{\Psi(\Gamma)}}_{\rm lc}.
\en 
Hereafter, $ \av{\cdots}_{\rm lc}$
denotes the average over $ P_{\rm lc}(\Gamma)$. 
Then, using  Eqs.(111) and (114), we  rewrite  Eq.(94) as 
\be 
d\Gamma_t
P_{\rm lc}(\Gamma_t) 
=d\Gamma_0 P_{\rm lc}(\Gamma_0) \exp[- {\cal D}(t)], 
\en 
As in Eq.(95),  phase-space integration of Eq.(116)  yields 
\be 
\av{e^{- {\cal D}(t)}}_{\rm lc}=
\int d{\Gamma}_0 P_{\rm lc} (\Gamma_0)
e^{- {\cal D}(t)}= 1,  
\en 
which yields 
${\bar{\cal D}}(t)= \int d\omega  
\omega f_{\rm lc}(\omega, t)>0$,  supporting   the second law of thermodynamics  in the bulk region.

\subsection{Steady-state distribution in nonlinear regime} 

  To seek  the steady-state distribution $P_{\rm st}(\Gamma)$, we 
start with the local-equilibrium one   $ P_{\rm lc}(\Gamma)$ 
at $t=0$ assuming a stepwise boundary-temperature change. 
The relaxation   $P_{\rm lc}\to P_{\rm st}$ 
should  take place  as microscopic events 
in a microscopic time in the preexisting temperature gradient, 
as was discussed  by Mori\cite{Mori1}. 
On the other hand, the relaxation from  $P_{\rm e}$ 
occurs slowly  on the time scale of $t_D$. 
Here, for the choice    $ P(\Gamma,0)= P_{\rm lc}(\Gamma)$, 
the  dissipation function is given by  
$ {\cal D}(t)$ in Eq.(112). In fact, from  Eqs.(110)-(112), 
time derivative of  Eq.(92) becomes  
\be 
\frac{d}{dt}\Big [ \beta{\cal H}(t)+\Psi(t) 
+ \sum_K \frac{\gamma}{2}\zeta_K(t)^2\Big]= 
{\Lambda }(t)+ {\dot{\cal D}}(t) ,
\en 
where ${\Psi}(t)={\Psi}(\Gamma_t)$. 
This again leads to Eq.(116).  Here,   
 use is made of  the approximation in 
 the second line of  Eq.(12), which is valid for large $H$.

The counterparts of  Eqs.(96) and (108) are given by 
\bea 
&&
P(\Gamma,t )= P_{\rm lc}(\Gamma) \exp[ {\cal D}(t)],\\
&&
P(\Gamma^*,t )= P_{\rm lc}(\Gamma) \exp[- {\cal D}(t)].
\ena 
As the counterpart of Eq.(109),  Eq.(120) yields  
\be
{\bar J}^z_h(z,t) 
=- \av{ J_h^z({\bi r},0)e^{-{\cal D}(t)}  }_{\rm ls}, 
\en 
 which  tends to  $ -\lambda{\cal T}$ 
homogeneously in a short time 
 in contrast to  ${\bar J}^z_h(z,t)$  in Eq.(109).
Thus, we find   
\bea 
&&\hspace{-1cm}P_{\rm st}(\Gamma)=   P_{\rm lc}(\Gamma)   
  \exp[{{\cal D}(t_{\rm lc})}] ,\\
&&\hspace{-1cm} P_{\rm st}(\Gamma^*)= 
  P_{\rm lc}(\Gamma) 
  \exp[{-{\cal D}(t_{\rm lc})}],
\ena 
where $t_{\rm lc}$ is taken to be    longer than 
the relaxation time of 
$P_{\rm lc}\to P_{\rm st}$. In Eqs.(119) and (121) we  use  backward 
 trajectories $\Gamma_{-s}$ 
($0<s<t)$  with $\Gamma_0=\Gamma$ 
and 
\be  
 {\cal D}(t)= -\gamma_{\rm a} 
\int_0^t ds {\cal G}(-s),
\en 
 where  we change  
$  {\cal G} (s)$  in Eq.(112) to $ {\cal G} (t-s)$  
 and   shift the time origin  by $-t$. 
 In Eqs.(120) and (123), we  use  
forward trajectories  $\Gamma_s$ ($0<s<t)$  with $\Gamma_0= \Gamma$.

We can also 
assume  that  $P_{\rm st}(\Gamma)$ is given by the time average 
of  $P(\Gamma,t)$ in a time  interval 
with width longer than $t_{\rm lc}$. Then, 
 the Laplace transformation $\int_0^\infty dt e^{-\epsilon t}
 P(\Gamma,t)$  tends to $P_{\rm st}(\Gamma)/\epsilon$ 
for  $0<\epsilon \ll t_{\rm lc}^{-1}$. 
With the aid of    $\epsilon e^{-\epsilon t} = -d(e^{-\epsilon})/dt$, 
we   obtain 
\bea 
&& \hspace{-10mm} 
{P_{\rm st}(\Gamma)}= {P_{\rm lc}(\Gamma) }
\Big[ 1-  
\int_0^\infty\hspace{-1mm}  dt~e^{-\epsilon t}~
  {\cal G}(-t) \gamma_{\rm a} 
e^{{\cal D}(t)}\Big],\\
&& \hspace{-10mm} 
{P_{\rm st}(\Gamma^*)}= {P_{\rm lc}(\Gamma) }
\Big[ 1+ 
\int_0^\infty\hspace{-1mm}  dt~e^{-\epsilon t}~  {\cal G}(t) \gamma_{\rm a} 
e^{-{\cal D}(t)}\Big],
\ena 
which readily give   the linear forms (69) and (70).

In the early literature\cite{McL,Zu,Gunton,Oppen}, 
the steady-state distribution  was  
 expressed  in the form of Eq.(122) for simple fluids, where   
 $t_{\rm lc}$ was pushed to $\infty$. 
 In particular, 
Kawasaki and Gunton\cite{Gunton}  
studied   sheared steady states with  $P_{\rm st} /P_{\rm lc} =
 \exp[ -\beta \int_0^\infty dt 
\int d{\bi r}\Pi_{zx}({\bi r},-t)  {\dot \gamma}]$,  where 
  $\Pi_{zx}$ is  the shear stress    and $\dot\gamma$ 
is the  shear rate. They  calculated  the nonlinear 
shear viscosity due to the mode-coupling effect, where the 
life times of the  shear  modes   are cut off by applied 
shear at long wavelengths\cite{Onukibook}.

\section{Summary and Remarks}

We have presented a microscopic theory of applying
a heat flow  from thermostatted boundary walls 
and have  derived Green's expression 
for the thermal conductivity $\lambda$ in the bulk.
 Our theory is based on the surface-to-bulk connecting 
 relationship in Eqs.(13) and (25). 
We give only the boundary tempeatures 
and  do not assume a constant temperature gradient 
in the interior, so our method is applicable to
 any inhomogeneous systems.  We summarize our main results as follows.\\
(i) In Sec.II, we have explained our system composed of unbound 
particles in the cell, those bound to the boundary layers, 
and  thermostats attached to the layers.  We have 
introduced  the heat flows ${\cal I}_K$ from the bound particles  
to the unbound ones in Eq.(11) and those  ${\cal J}_K$ from the thermostats 
to the particles in Eq.(23). They have 
  bulk  expressions in Eqs.(13) and (25) 
as key relations. \\
(ii) In Sec.III, we have derived linear response 
relations to small boundary temperature changes $\delta T_K(t)$ 
from   the Liouville equation. 
Their surface expressions have been given in Eqs.(41) and (42) 
in terms of the time-correlation functions 
$\av{{\cal B}(t) {\cal J}_K(0)}_{\rm e}$. Using  Eq.(25), 
we have also obtained  the  bulk response expressions 
composed of  local-equilibrium and dissipative 
parts. We have also calculated 
 the nonequilibrium  average of the Evans-Searles 
dissipation function $\Omega(t)$\cite{W,FD,Evans1,Searles1} generally 
in terms of the hydrodynamic variables in Eqs.(56) and (61) 
and explicitly for a stepwise  temperature change in Eq.(62). 
\\ 
(iii) In Sec.IV, we have examined 
the phase-space distribution $P(\Gamma,t)$ in the nonlinear regime. 
 First, we have summarized 
salient results when the initial distribution  $P(\Gamma,0)$ 
is the equilibrium one. 
In particular, for $\delta T_{\rm top}(t)= \delta T_{\rm bot}(t)$, 
we have obtained simple results including Eqs.(101) and (102). 
Furthermore, using Eq.(25), we 
have introduced the local-equilibrium distribution 
$P_{\rm lc}(\Gamma)$ in Eq.(114).  In the case  
$P(\Gamma,0)= P_{\rm lc}(\Gamma)$, 
 we have obtained the  steady-state one  $P_{\rm st}(\Gamma)$ 
 in the McLennan-Zubarev form\cite{Zu,McL} in Eqs.(122) and (123). 
\\
(iv)  We have  examined   the linear relaxations 
 of the hydrodynamic variables in Appendix A and 
 the dissipattion function in in Appendix B  
after a boundary temperature change. These results 
  enable us to calculate the time-correlation functions  
$\av{{\cal B}(t) {\cal J}_K(0)}_{\rm e}$. 
 We have presented a theory of 
the long-range correlation 
in the canonical ensemble  in Appendix C, which 
leads to   the local-equilibrium 
response in steady states. In Appendix D, 
we have calculated $\Omega(t)$ including $T$-dependence of $\tau_{\rm h}$.

We make some remarks as follows.
(1) Though we have used  Nos\'e-Hoover 
thermostats\cite{Nose,Hoover}, our results should be independent of the 
thermostating method on long-time scales ($\gg \tau_{\rm h}$). 
(2) We should examine the thermal boundary 
resistance at a solid-fluid interface in more 
detail. Its Green-Kubo type 
formula\cite{Barrat,Cast} has not yet been firmly established 
from our viewpoint. 
(3) It is of great interest to 
generalize our results to multi-component fluids, 
where the thermo-diffusion effect is crucial.
(4) Our scheme is applicable to systems in the presence 
of two-phase interfaces\cite{Hafs,Cast}  
and to mesoscopically heterogeneous systems 
such as glasses and  polycrystals. 
(5) In our previous paper on shear strains\cite{Kawasaki} 
we  examined only the linear response. 
We should further examine sheared states 
in the presence of thermostats.  
(6) Numerical study of 
   the relaxation  $P_{\rm lc}\to P_{\rm st}$ 
 should be informative, which is  easy particularly 
 for  one-dimensional systems. 

\acknowledgments{
I would like to thank 
Takeshi Kawasaki for valuable  discussions 
on the  strain effect in glasses leading  to this work. 
I am also indebted to  Hisao Hayakawa 
for useful correspondence on 
the fluctuation theorems.  
}

\vspace{2mm}
\noindent{\bf Appendix A: Thermal   relaxation after a 
boundary temperature change at a fixed volume }\\
\setcounter{equation}{0}
\renewcommand{\theequation}{A\arabic{equation}}

To examine slow dynamics in the film region  $0<z<H$, 
we  treat   averaged smooth   
quantities obeying the linearlized hydrodynamic 
equations on timescales much 
longer than the typical molecular time $t_{\rm m}$. 
We start with   the   heat conduction equation\cite{Landau-f,Kada}, 
\be 
nT \frac{\p}{\p t}\delta s =   \lambda \nabla_z^2\delta T, 
\en 
without significant  Kapitza resistance. 
Here,  $\delta s$ is the deviation of the entropy per particle 
 related to that  of  
 the temperature $\delta T$ and that of  the pressure  
 $\delta p$ by   
\be 
nT \delta s= C_p[ \delta T- (\p T/\p p)_s \delta p].  
\en  
Here, $nT\delta s$ is  the average  
deviation of ${\hat q}$ in Eq.(4)  and $\delta p$ is that 
of $\hat p$ in Eq.(8) slightly away from equilibrium.
They are related  to $\chi_{aK}(z,t)=
\av{{\hat a}({\bi r},t){\cal J}_K(0)}_{\rm e}$ 
for $\hat{a}= {\hat q}$ and $\hat p$ (see  Sec.IIIA). 
In the linear order, the temperature deviation $\delta T$ 
is defined by Eq.(A2). 
The average boundary heat fluxes    ${\cal I}_K(t) (\cong {\cal J}_K(t))$  
are  written as  
\be 
{\cal I}_{\rm top}(t)= A\lambda \delta T'(H,t),
 ~{\cal I}_{\rm bot}(t)= -A \lambda \delta T'(0,t),
\en 
where $A$ is the surface area and  $\delta T'(z,t)= \p \delta T/\p z$.

 Let   the boundary temperatures 
be changed  by constant $ \delta T_K$  
in a time range $[0,t_{\rm ex}]$ with 
 $ t_{\rm ex}<t_{\rm a}=H/c$. 
For example, we can assume the linear increase:
 $\delta T_K(t)/\delta T_K= t/t_{\rm ex}$ for $0<t<t_{\rm ex}$.

{\it Initial stage}. For $t_{\rm m}\ll  t < t_{\rm a}$, 
we treat  the thermal diffusion near the  walls.  
Along the $z$ axis,   
$\delta T(z,t) $ decreases  from $\delta T_{K}(t)$ at the boundaries 
and  decays  to 0 far from them 
 on the diffusion length   $\sqrt{Dt}$. Thus, 
for $0<t<t_{\rm a}$, ${\cal I}_K(t)$ and $\delta{\cal H}(t)$ increase as  
\bea 
&&{\cal I}_K(t)\sim A\lambda \delta T_K(t)/\sqrt{Dt},\\
&& \delta {\cal H}(t) 
\sim \sum_K A\lambda \delta T_K(t)\sqrt{t/D}.
\ena 
See Eq.(63) for  the  corresponding response functions.

{\it Intermediate and final stages}. 
For $t\gg t_{\rm a}$,     $\delta p$ is known to be 
homogenized after repeated sound traversals in the cell 
(the piston effect)\cite{Za,Ferrell,Miura,Onukibook}, where 
$\delta p$ is equal to 
the  space average of $ (\p p/\p T)_n\delta T+ 
(\p p/\p n)_T\delta n$. Thus,   
\be 
\delta p(t)= (\p p/\p T)_n \av{{\delta { T}}}_{\rm sp}(t),
\en 
where    $\av{\delta{T}}_{\rm sp}= \int_0^H dz \delta T(z,t)/H$.  As a result, 
the  temperature increases  by $(\p T/\p p)_s\delta p$   
adiabatically  throughout the cell. 
This effect  is amplified  near the criticality, where the ratio  
$
 (\p T/\p p)_s/(\p T/\p p)_n=1-\gamma_{\rm s}^{-1}
$ 
is close to 1 with  $\gamma_{\rm s}=C_p/C_V\gg 1$.
The inhomogeneous part of   $\delta T$ near the walls 
is  governed by the thermal diffusion.

For $t\gg t_{\rm a}$,  we can use Eq.(A6) to  
 calculate  the Laplace transformation (LT) of $\delta T(z,t)$ defined by  
\be
F_T(z,\omega)= \int_0^\infty dt e^{-\omega t}\delta T(z,t),  
\en 
which is valid  for $\omega \ll t_{\rm a}^{-1}$. Some calculations 
 give\cite{Ferrell}    
\bea 
&&\hspace{-1cm}  F_T=\Big [1 
 +\frac{\cosh v- \cosh u}{\varphi(u)}
\Big]  \frac{\delta T_{\rm m}}{\omega} 
 + \frac{H\sinh v}{2\sinh u}\cdot\frac{{\cal T}}{\omega},
\ena 
where $\delta T_{\rm m}$ and ${\cal T}$ are given in 
Eqs.(35) and (36) and   
\bea 
&& u=H ({\omega/4D})^{1/2}= (t_D \omega)^{1/2}, \\
&& v=u (2z/H-1),\\   
&&\hspace{-1cm}
\varphi(u) = \cosh u+(\gamma_{\rm s}-1)u^{-1} \sinh u.
\ena 
Here, $F_T(z,\omega)$ 
 depends on $z$ through $v$ with $v=\pm u$ at $z=0$ and $ H$. 
From Eq.(A8) its space average    is given by    
\be 
\av{F_T}_{\rm sp}= \ppp{T}{p}{n} F_p 
= \frac{\gamma_{\rm s} \sinh u}{u\varphi(u)}
\cdot\frac{\delta T_{\rm m}}{\omega} ,
\en
where $F_p(\omega)$ is the LT of $\delta p(t)$ in Eq.(A6).  
The LTs:   $F_K(\omega)=\int_0^\infty dt e^{-\omega t}
{\cal I}_K(t)$ are also given by    
\be 
 F_K(\omega) = A\lambda t_D\Big[
 \frac{2\sinh u}{Hu\varphi(u)}\delta T_{\rm m} \pm 
\frac{{\cal T}}{u\tanh u}\Big], 
\en
where $+$ is for $K=$top and 
$-$ is for $K=$bot.  For $\gamma_{\rm s}\gg 1$, 
we have near-critical behavior 
$\varphi\cong (\gamma_{\rm s}-1)/u\gg 1$ for 
$1\ll u\ll \gamma_{\rm s}$ (see Eqs.(B7) and (B8)).

We remark the following. 
(i) We find   $ F_T(z,\omega)\to \delta  {\bar T}(z)/\omega$ 
as $\omega \to 0$ 
and $\delta T(z,t)  \to  \delta {\bar T}(z)$  as $t \to \infty$, where 
 $\delta {\bar T}(z)$ is given in Eq.(54) in the steady state.  
 (ii) For $u\gg 1$, we have 
$ \omega \av{F_T}_{\rm sp}\cong 
\delta T_{\rm m} \gamma_{\rm s} /(u+ \gamma_{\rm s}-1)$. 
 Thus, for   $\gamma_{\rm s}\gg 1$,   we have 
$\av{\delta{T}}_{\rm sp}(t) \cong \delta T_{\rm m}$ 
 for   $t \gs t_1$, where  
\be  
t_1= ( \gamma_{\rm s}-1)^{-2}H^2/4D \ll t_D. 
\en 
in particular,  for ${\cal T}=0$, Thus, thermalization occurs at $t\sim t_1$. 
We assume    $t_1\gg t_{\rm a}$, which  
holds for $H\gg \gamma_{\rm s}^2 D/c$, while Eq.(A8) 
holds even in the reverse case $t_1 \ls t_{\rm a}$. 
(iii) The second term in Eq.(A8) 
 gives rise to  the temperature relaxation 
  approaching  the linear profile 
 ${\cal T}(z-H/2)$ diffusively on the timescale of $t_D$.

\vspace{2mm}
\noindent{\bf Appendix B: Dissipation function 
in  hydrodynamic regime }\\
\setcounter{equation}{0}
\renewcommand{\theequation}{B\arabic{equation}}

In this appendix, we first 
 derive Eq.(56).  
From   Eqs.(A1)-(A3)  the hydrodynamic 
average   of the dissipation function in Eq.(50)   is written as     
\bea 
&&\hspace{-6mm} \frac{ {\bar{\Omega}}(t) }{A} 
= \frac{\lambda}{k_BT^2} \int_0^t \hspace{-1mm}dt \Big[ (\delta T 
 \delta T')_{z=H} - (\delta T \delta T')_{z=0}\Big] \nonumber\\
&&\hspace{-3mm}
 = \frac{1}{k_BT} \int_0^t\hspace{-1mm} dt \int_0^H \hspace{-1mm}dz 
\Big[ n \frac{\p \delta { s} }{\p t} \delta T 
 +\frac{\lambda}{T}  ( \delta T')^2\Big]. 
\ena 
To derive the second line,  we change 
 the integrand  in the first line  to  
$\int_0^H dz [\delta T \delta T''+ (\delta T')^2]$ 
and use   Eq.(A1). 
We also calculate the time derivative of $(\Delta S)_2$ in Eq.(57) as  
\be 
\frac{d}{dt} \int d{\bi r}(\Delta S)_2= - 
\int d{\bi r}\Big[ \frac{n}{T}\frac{\p \delta s}{\p t}\delta T+ 
\frac{\p\delta n}{\p t}\frac{\delta p}{nT}\Big],
\en 
where  $C_V\delta T= nT\delta s+T(\p p/\p T)_n\delta n/n$. Here,  
the   space integral of 
$({\p\delta n}/{\p t}){\delta p}/{n}$ 
is changed to that of 
 $ {\bi v}\cdot\nabla \delta p
=  
{\bi v}\cdot[-\rho   {\dot{\bi v}}+
\nabla\cdot{{\asigma}}_{\rm v}]$  from 
$\p (\delta n)/\p t= -n\nabla\cdot{\bi v}$, where 
${{\asigma}}_{\rm v}$ is  the viscous stress tensor.  
Thus, we are led to   Eq.(56).

Second, we explicitly calculate ${\bar\Omega}(t)$. 
In  the initial stage $0<t<t_{\rm a}$, 
the bottom and top disturbed  regions are 
separated, so    Eq.(A4) yields  
\be
{{\bar{\Omega}}(t)}/{A}\sim \lambda \sqrt{t/D} 
 \sum_K [\delta T_K(t)]^2/k_BT^2.
\en
For $t\gg t_{\rm a}$,   Eq,(A8) gives  the LT of 
${\bar{\Omega}}(t)$  in the form,  
\be 
{F_{\Omega}(\omega)}
=\frac{V\lambda t_D}{k_BT^2\omega}\Big [\frac{4\sinh u}{ u\varphi(u)} 
\cdot\frac{(\delta T_{\rm m})^2}{H^2}+ \frac{{\cal T}^2}{ 
u\tanh u}\Big], 
 \en 
whose inverse LT yields Eq.(62). Using $\tau=t/t_D$ we have 
\bea 
&&\hspace{-1cm}
\psi_1(\tau)= \sum_{\ell \ge 1} 
\frac{2\gamma_{\rm s}}{a_\ell^2+ \gamma_{\rm s} (\gamma_{\rm s}-1)}
\Big[1- \exp(-a_\ell^2 \tau)\Big],\\
&&\hspace{-5mm}
\psi_2(\tau)= \sum_{\ell \ge 1} \frac{2}{\pi^2\ell^2}\Big
[1- \exp(-\pi^2\ell^2 \tau)\Big]. 
\ena 
In Eq.(B5),   $a_\ell$ ($\ell\ge 1$) are 
the solutions of $\tan a_\ell= - a_\ell/(\gamma_{\rm s}-1)$ 
with  $\ell-1/2<a_\ell/\pi< \ell$.
The  inverse LTs  of $t_D\gamma_{\rm s} {\sinh u}/{u\varphi(u)}$ and 
$t_D/({u\tanh u})-1/\omega$ are  $\psi_1'=d\psi_1/d\tau$ and 
 $\psi_2'=d\psi_2/d\tau$, respectively,   
which indicate  $\psi_1(\infty)=1$  and  $\psi_2(\infty)=1/3$.
For $\gamma_{\rm s}\gg 1$, $\psi_1$ depends on $\gamma_{\rm s}$ in a singular 
manner as\cite{Ferrell} 
\bea  
&&\hspace{-1.2cm}
\psi_1(\tau)\cong 2(\gamma_{\rm s}-1) (\tau/\pi)^{1/2} \quad (t< t_1 )\\
&&\hspace{-2mm}
\cong 1- (\gamma_{\rm s}-1)^{-1}/ (\pi \tau)^{1/2} \quad (t_1< t< t_D ).
\ena 
Thus, $\psi_1$ appraoches to 1 for 
$t\sim t_1$ (the piston effect).

\vspace{2mm}
\noindent{\bf Appendix C: Long-range correlation 
in finite systems in  canonical ensemble}\\
\setcounter{equation}{0}
\renewcommand{\theequation}{C\arabic{equation}}
  
We examine  the long-range correlations  
in a finite system  in the $T$-$N$-$V$ canonical 
ensemble\cite{Lebo,Vlugt,Kremer,Rogers}, which are absent in 
the $T$-$\mu$-$V$ grand canonical 
ensemble.  We treat macroscopic one-component fluids, but 
extension to multi-component fluids\cite{Rogers,Kremer,Vlugt} 
 is straightforward. The average 
$\av{\cdots}_{\rm e}$ is taken at fixed $V$,  $T$, and $N=nV$.

In the canonical ensemble, 
the density correlation function $g({\bi r},{\bi r}')= 
\av{\delta {\hat n}({\bi r}) \delta {\hat n}({\bi r}')}_{\rm e}$ 
in the bulk  is  of the  form, 
\be
g({\bi r},{\bi r}')=
n\delta({\bi r} -{\bi r}')+  n^2g(|{\bi r} -{\bi r}'|)
 - k_BT n^2\kappa_T/V,
\en
where  $g(r)$ is the pair correlation function   
and  $\kappa_T$ is the isothermal compressibility. Space 
integration of  Eq.(C1) in the  
cell vanishes from $1+ n\int d{\bi r}g(r) = k_BT n\kappa_T$. 

For any 
density variables $\hat{a}({\bi r})$ and $\hat{b}({\bi r})$, we 
consider their deviations  $\delta\hat{a}({\bi r})
=\hat{a}({\bi r})-a$ and $\delta\hat{b}({\bi r})= 
\hat{b}({\bi r})-b$, where   the averages 
 $a= \av{\hat{a}}_{\rm e}$  and 
$b= \av{\hat{b}}_{\rm e}$ are    thermodynamic quantities. 
In the canonical ensemble, we have 
\be 
\av{\delta {\hat a}({\bi r}) \delta {\hat b}({\bi r}')}_{\rm e}=
g_{ab}({\bi r} -{\bi r}') - k_BT n^2 \kappa_T D_{ab}/V, 
\en 
where $g_{ab}({\bi r} -{\bi r}') $ is a short-ranged function and 
\be 
 D_{ab}=  \ppp{a}{n}{T}\ppp{b}{n}{T}= 
 \frac{1}{n\kappa_T}  \ppp{a}{p}{T}\ppp{b}{n}{T}.
\en 
Here, if $\hat{b}=  \hQ$, 
  Eq.(79) follows from   
$   n^2(\p s/\p n)_T= -(\p p/\p T)_n $. 
 If ${\hat b}={\hat p}$, 
the second term of Eq.(C2) is $- nk_BT (\p a/\p n)_T/V$. 
If  $\hat{b}$ is  the temperature fluctuation 
$\delta{\hat T}\equiv 
C_V^{-1}\delta{ \hat q}+ (\p T/\p n)_s \delta {\hat n}$,  
we have $D_{ab}=0$.

A general theory of Lebowitz {\it et al.}\cite{Lebo}  
can yield Eq.(C2), but 
we here present its  simple derivation. We divide the cell  
into  regions $A$ and $B$ with volumes $V_A$ and $V_B=V-V_A$,  
 where  $\bi r$ is in  A and ${\bi r}'$ is in  B far from the boundaries. 
 We examine  the fluctuation of the particle number 
 $N_A$ in A  around ${\bar N}_A = nV_A$. Then, 
for each $N_A$, the conditional average of 
$\hat{a}({\bi r})$  is $a(T, N_A/V_A)$, so    
\be 
\delta\hat{a}({\bi r})= 
a(T, N_A/V_A)- a(T,n)\cong \ppp{a}{n}{T} \frac{\delta N_A}{V_A},
\en  
 where $|\delta N_A | \ll {\bar N}_A$.  In B, 
   we have  $N_B= {\bar N}_B-\delta N_A$ with ${\bar N}_B = nV_B$ so that 
$\delta\hat{b}({\bi r}')= - (\p b/\p n)_T \delta N_A/V_B$.  
Thus, 
\be 
\av{\delta {\hat a}({\bi r}) \delta {\hat b}({\bi r}')}_{\rm e}=
- D_{ab} 
\sum_{ \delta N_A} P( N_A) ({\delta N_A })^2/{V_A V_B},
\en 
where   $P(N_A)$ is the distribution  of   $N_A$ proportional to 
 $\exp[ - F(T,N_A)/k_BT - F(T, N-N_A)/k_BT]$, 
where $F$ is the  Helmholtz free energy. Up to second order, we   find 
\be 
P( N_A)\propto  \exp[ - 
(V/2k_BT \kappa_T {\bar N}_A {\bar N}_B)(\delta N_A)^2].  
\en  
 Using this Gaussian distribution 
we sum over $\delta N_A$ in   Eq.(C5) 
to obtain  the second term in Eq.(C2).

Finally, we 
 write  the space integral of the first term in Eq.(C2)  as 
$\av{{\hat a}:{\hat b}}\equiv  \int d{\bi r}g_{ab}({\bi r})$ in the 
grand canonical ensemble.  Then,   we have 
$ 
 \av{{\hat a}:{\hat n}}
= nk_BT ({\p a}/{\p p})_{T}$ and $ \av{{\hat a}:{\hQ}}
= k_BT^2 ({\p a}/{\p T})_{p}$\cite{Onukibook}.
We can now rewrite the second term in   Eq.(C2)   as 
$
-{
\av{{\hat a}:{\hat n}}\av{{\hat b}:{\hat n}}}/{V\av{{\hat n}:{\hat n}}}.
$. 

\vspace{2mm}
\noindent{\bf Appendix D: Dissipation function including temperature-dependence of $\tau_{\rm h}$ }\\
\setcounter{equation}{0}
\renewcommand{\theequation}{D\arabic{equation}}

We here derive the  dissipation function $\Omega(t)$ 
for  $T$-dependent  $\tau_{\rm h}$ in the nonlinear theory,  where  
 $\tau_{\rm h}=\tau_{\rm h}(T_K(t))$ 
depends on $t$ and $K$. 
From  Eq.(17), we have 
\be 
 {\dot \zeta}_K(t) = [2E_K(t) - dM k_BT_K(t)]/Q_K(t), 
\en 
where  $
Q_K(t)= [\tau_{\rm h}((T_K(t))]^2 dM k_BT_K(t)$. 
 Evans and Searles\cite{Evans1,Searles1,W} treated 
 $Q_K(t)$ as  a constant  $Q$. 

 For  $P(\Gamma,0)= P_{\rm e}(\Gamma)$,  
  the first line of Eq.(93) gives     
\bea
&& {\dot\Omega}(t) =  2\sum_K\delta\beta_K(t) E_K(t)\zeta_K(t)  \nonumber\\
&& -  \sum_K \Big
[\beta_K(t) {Q_K(t)}  -\beta{Q_K(0)} \Big] 
   \zeta_K(t){\dot\zeta}_K(t)  .
\ena 
The first term  yields     Eq.(50) from Eq.(23), while 
the second term arises from   $T$-dependence of 
$\tau_{\rm h}$ from    $ \beta_K(t) {Q_K(t)} =dM [\tau_{\rm h}(T_K(t))]^2$. 
In particular, if  $Q_K(t)=Q_K(0)$ or if $\tau_{\rm h}^2\propto T^{-1}$, 
  Eqs.(D1) snd (D2) yield  
\be 
{\dot{\Omega}} (t)= - dM  \sum_K  \zeta_K(t) \delta T_K(t)/T,  
\en  
whose time-integration  gives the  Evans-Searles 
$\Omega(t)$ for heat conduction\cite{Evans1,Searles1,W}. 
 However, for any $T$-dependence of $\tau_{\rm h}$, 
the  difference 
between our $\Omega(t)$ and theirs is 
negligible for $t\gg \tau_{\rm h}$, since  
the second term in Eq.(D2) 
is proportional to $d{\zeta}_K(t)^2/dt$. 
 In the linear regime,  Eq.(D2) gives  
Eq.(48)  from   $\tau_{\rm h}(T_K)-\tau_{\rm h}(T) \cong 
(\p \tau_{\rm h}/\p T) \delta T_K$. In particular, 
we have    ${\cal J}_K'(t)= -dMk_BT  \zeta_K(t)$ 
for  $\tau_{\rm h}^2\propto T^{-1}$.

\end{document}